\documentclass[]{article}
\usepackage[marginclue,footnote,silent,draft]{fixme}
\usepackage{amsmath}
\usepackage{amssymb}
\usepackage{graphbox}
\usepackage{pbox}
\usepackage{lipsum} 
\usepackage[toc,page]{appendix}
\usepackage{siunitx}
\usepackage{multirow}
\usepackage{makecell}
\usepackage{enumitem}
\usepackage[breaklinks=true]{hyperref}
\usepackage{breakcites}
\usepackage{natbib}
\usepackage{amssymb}
\usepackage{latexsym}
\usepackage{graphicx}
\usepackage{subfigure}
\usepackage[marginclue,footnote,silent,draft]{fixme}
\usepackage{multirow}
\usepackage{amsmath}
\usepackage{amssymb}
\usepackage{url}
\usepackage{hyperref}
\usepackage{pgf,tikz}
\usepackage{tikz}
\usetikzlibrary{intersections}
\usetikzlibrary{spy}
\usetikzlibrary{arrows,automata}
\usepackage{xcolor}
\usepackage{ragged2e}
\usepackage[export]{adjustbox}
\usepackage[marginclue,footnote,silent,draft]{fixme}
\usepackage{titlesec}
\usepackage{amsfonts}
\usepackage{algorithmic}
\usepackage{textcomp}

\definecolor{bluSezione}{RGB}{19,138,218}
\graphicspath{ {images/} }
\begin{document}
\title{A Universal Deep Learning Framework for Real-Time Denoising of Ultrasound Images}
\author{Simone Cammarasana, Paolo Nicolardi, Giuseppe Patan\`e\\
CNR-IMATI~$\&$ ESAOTE SpA}
\maketitle
\begin{abstract}
Ultrasound images are widespread in medical diagnosis for muscle-skeletal, cardiac, and obstetrical diseases, due to the efficiency and non-invasiveness of the acquisition methodology. However, ultrasound acquisition introduces noise in the signal, which corrupts the resulting image and affects further processing steps, e.g., segmentation and quantitative analysis. We define a novel deep learning framework for the real-time denoising of ultrasound images. Firstly, we compare state-of-the-art methods for denoising (e.g., spectral, low-rank methods) and select WNNM (\emph{Weighted Nuclear Norm Minimisation}) as the best denoising in terms of accuracy, preservation of anatomical features, and edge enhancement. Then, we propose a tuned version of WNNM (\emph{tuned-WNNM}) that improves the quality of the denoised images and extends its applicability to ultrasound images. Through a deep learning framework, the tuned-WNNM qualitatively and quantitatively replicates WNNM results in real-time. Finally, our approach is general in terms of its building blocks and parameters of the deep learning and high-performance computing framework; in fact, we can select different denoising algorithms and deep learning architectures.
\end{abstract}
\section{Introduction}
\emph{Ultrasound imaging} uses high-frequency sound waves to visualise soft tissues, such as internal organs, and support medical diagnosis for muscle-skeletal, cardiac, and obstetrical diseases, due to the efficiency and non-invasiveness of the acquisition methodology. Ultrasonic sound waves are reflected off from different layers of body tissues. The main issues of the ultrasound techniques are a significant loss of information during the reconstruction of the signal, the dependency of the signal from the direction of acquisition, an underlying noise that corrupts the image and significantly affects the evaluation of the morphology of the anatomical district. In this context, the denoising of ultrasound images is relevant both for post-processing and visual evaluation by the physician.

\begin{figure*}
\includegraphics[width=0.99\textwidth,left]{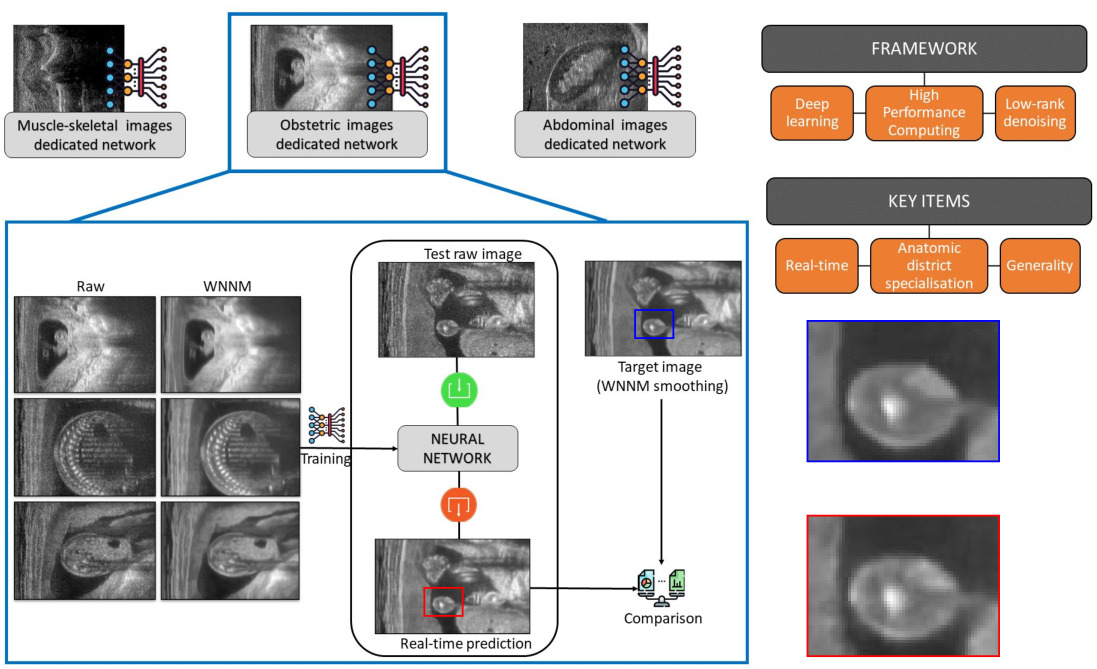}

\caption{Pipeline of the proposed framework, with a magnification of prediction (red) and target (blue) images (right side).
Our framework applies deep learning and HPC to learn and replicate the denoising results of state-of-the-art low-rank method (i.e., tuned-WNNM), in real-time and with a specialisation to anatomic districts.\label{FIG:TEASER}}
\end{figure*}

Our main goal is the definition of a novel deep learning framework for the real-time denoising of ultrasound images (Fig.~\ref{FIG:TEASER}). After the design of a training data set, composed of raw images and the corresponding denoised images, we train a neural network that replicates the denoising results. Then, the real-time denoising is achieved through the prediction of the trained network. The proposed framework combines three elements: \emph{low-rank denoising}, \emph{deep learning}, and \emph{high-performance computing} (HPC, for short). 
 
We select WNNM-\emph{Weighted Nuclear Norm Minimisation}~\cite{gu2014weighted} as the best denoising method, which is then specialised to ultrasound images as a ``new'' \emph{tuned-WNNM} denoising, by tuning its parameters. The choice of WNNM is based on a qualitative and quantitative comparison of five denoising methods, i.e., WNNM, SAR-BM3D - \emph{SAR Block-Matching 3D}~\cite{parrilli2011nonlocal}, BM-CNN - \emph{Block Matching Convolutional Neural Network}~\cite{ahn2017block}, NCSR - \emph{Non-Locally Centralised Sparse Representation} method~\cite{dong2012nonlocally}, PCA-BM3D \emph{Principal Component Analysis Block Matching 3D}~\cite{dabov:inria-00369582}) belonging to the spectral, low-rank, and deep learning classes~(Sect.~\ref{SEC:RELWORK}).

To achieve a \emph{real-time denoising of ultrasound images}, we propose a deep learning framework that is based on the learning of the tuned-WNNM and HPC tools (Sect.~\ref{sec:METHOD}). The training is performed offline and can be further improved with new data, a-priori information on the input images or the anatomical district, and denoised images selected after experts' validation. Through our framework, the execution time of the denoising only depends on the network prediction, which is achieved in real-time on standard medical hardware.
 
As the main contribution, the proposed denoising of ultrasound images runs in real-time and is general in terms of the input data, in terms of resolution of the input images (e.g., isotropic, anisotropic images), acquisition methodology, anatomical district, noise (e.g., speckle or Gaussian noise), and the dimension of the images (i.e., 2D, 3D images). Our approach is also general in terms of the building blocks and parameters of the deep learning framework; in fact, we can select different denoising algorithms (e.g., WNNM, SARBM3D) and deep learning architectures (e.g., Pix2Pix, VGG19).
 
As experimental validation (Sect.~\ref{sec:RESULTS}), we perform a quantitative (e.g., PSNR, SSIM) and qualitative evaluation of the selected denoising methods on ultrasound images acquired from different anatomical (e.g., muscle-skeletal, obstetric, and abdominal) districts. Then, the results of the deep learning and HPC frameworks are quantitatively and qualitatively analysed on a large collection of ultrasound images. The industrial requirement of real-time denoising is verified in terms of the execution time of the network prediction on GPU-based hardware installed in commercial ultrasound machines. Finally, we discuss main results (Sect.~\ref{SEC:DISCUSSION}), conclusions, and future work (Sect.~\ref{SEC:CONCLUSIONS}). 

\section{Related work\label{SEC:RELWORK}}
We review previous work on image denoising~(Sect.~\ref{sec:IMAGE-despeckling}) and deep learning methods for denoising and regression~(Sect.~\ref{sec:DEEP-despeckling}).
\subsection{Image denoising\label{sec:IMAGE-despeckling}}

\paragraph{Non-local methods} The \emph{Non-Local Means} (NLM) denoising~\cite{buades2005non} uses the patterns' redundancy of the input image, and each patch is restored with a weighted average of all the other patches, where each weight is proportional to the similarity among the patches. The \emph{Bayesian non-local mean filter}~\cite{kervrann2007bayesian} improves the NLM with the introduction of a Bayesian estimator as distance measure among the patches, which allows the user to better determine the amount of denoising by the noise variance of the patch. The anisotropic neighbourhood in NLM~\cite{maleki2013anisotropic} uses image gradient to estimate the edge orientation and then adapts the patches to match the local edges. The characterisation of the patches through a redundancy index~\cite{MEI2020105670CMBP} improves the self-similarity computation among patches. The improvement on the structure of the search window is achieved through the computation of an optimal search window for each pixel~\cite{verma2017adaptive}, according to the denoising degree of the related patch.

\paragraph{Anisotropic methods}
The denoised image is computed as the solution to an anisotropic diffusion equation~\cite{perona1990scale,Patane:2015SMOOTH}, where the gradient of the image guides the diffusion process. The variant~\cite{yu2002speckle} exploits the Lee~\cite{4766994} and Forst~\cite{frost1982model} filters, which are edge-sensitive to speckle noise. An improvement of the previous results~\cite{aja2006estimation} is achieved by applying the Kuan filter~\cite{kuan1985adaptive} in the diffusion equation and revising the selection of the neighbourhood used for the estimation of the statistical parameters. The anisotropic method introduces a class of fractional-order anisotropic diffusion equations~\cite{bai2007fractional}, using the Fourier transform to compute the fractional derivatives, and the discrete Fourier transform to compute the fractional-order differences.
 
\paragraph{Spectral denoising}
Denoising based on \emph{spectral decomposition} transforms a signal into its spectral domain and exploits the sparsity of the transformed signal to remove noise, through a threshold operation. Several transformations have been applied to image denoising, such as Wavelets~\cite{mihcak1999low,chang2000adaptive,portilla2003image,liu2017efficient}, Curvelets~\cite{starck2002curvelet}, Contourlets~\cite{da2006nonsubsampled}, and Shearlets~\cite{yang2014image}. To reconstruct the denoised image, the \emph{3D block-matching}~\cite{dabov2006image} computes and stacks similar patches through NLM; each stack is transformed into its spectral domain with wavelet decomposition, denoised through a hard/soft threshold, and reconstructed in the space domain. The denoised patches are aggregated by a collaborative filter. The \emph{synthetic aperture radar block matching 3D} (SAR-BM3D)~\cite{parrilli2011nonlocal} introduces a speckle-based variant of 3D block matching; the similarity among the patches is computed by considering the probability distribution of the speckle noise as a distance metric. Furthermore, the hard/soft threshold of the wavelet transformed signal is substituted by a Local Linear Minimum Mean Square Error (LLMMSE) filter. The \emph{principal component analysis block matching 3D} (PCA-BM3D)~\cite{dabov:inria-00369582} improves the stacking operation of 3D block-matching by using shape-adaptive neighbourhoods, which enable its local adaptability to image features. The 3D transformation of each stack to the spectral domain is performed through the PCA~\cite{wold1987principal} and an orthogonal 1D transformation in the third dimension.

\paragraph{Low-rank methods}
Low-rank approximation computes the denoised image as the solution to a weighted minimisation problem, whose cost function is the Frobenius norm~\cite{srebro2003weighted, cammarasana2021universal, Patane:2015SMOOTH} or the~$\ell_{1}$ norm~\cite{eriksson2010efficient}, between the input and the target images. The relation between local and non-local information~\cite{dong2012nonlocal} allows us to estimate signal variances, by interpreting the \emph{Singular Value Decomposition} (SVD, for short) through a bilateral variance estimation. In~\cite{rajwade2012image}, a high-order SVD is applied to 3D blocks, and the denoised image is achieved with hard thresholding of the decomposed signal.
The \emph{Weighted Nuclear Norm Minimisation} (WNNM)~\cite{gu2014weighted} computes the stacks as in the 3D block-matching method, performs the SVD on the stacks and applies a weighted threshold to the singular values, where higher weights correspond to lower singular values, which capture the noisy component of the image. The \emph{collaborative filtering} of WNNM for the aggregation of the denoised patches is performed as in the 3D block-matching method. The weighted nuclear norm and the histogram preservation~\cite{zhang2018structure} are combined in a single constrained optimization problem, which is solved through the alternating direction method of multipliers~\cite{10.1561/2200000016}. The WNNM is extended to image deblurring with several types of noise~\cite{ma2017low}.

\paragraph{External learning}
The K-SVD algorithm~\cite{aharon2006k} represents the signal as a linear combination of an over-complete dictionary of atoms, which are iteratively updated through the SVD of the representation error to better fit the data. A \emph{learned simultaneous sparse coding} method~\cite{mairal2009non} integrates sparse dictionary learning with non-local self-similarities of natural images. The \emph{non-locally centralised sparse representation} (NCSR)~\cite{dong2012nonlocally} exploits the non-local redundancies, combined with local sparsity properties, to estimate the coefficients of the sparse representation of the input image. The dictionary is learned by clustering the patches of the image into~$K$ clusters through the K-means~\cite{macqueen1967some} method and then learning a PCA sub-dictionary for each cluster. This method has been further improved in~\cite{xu2017fast} with a fast version based on a pre-learned dictionary and achieving an improvement of the computational efficiency. The structured sparse model selection over a family of learned orthogonal bases~\cite{ma2016image} is applied to the deblurring of images with Gaussian noise.

\subsection{Deep learning for denoising and regression\label{sec:DEEP-despeckling}}
\paragraph{Deep learning methods for denoising}
In the \emph{Noise2Noise} algorithm~\cite{lehtinen2018noise2noise}, the network learns to denoise images only considering the noisy data, without any knowledge of the ground-truth. The \emph{Noise2Void} algorithm~\cite{krull2019noise2void} further expands this idea, and it does not require couples of noisy images for the training. This approach is relevant in biomedical fields, where there are no ground-truth images. The \emph{Noise2Self} method~\cite{batson2019noise2self} proposes a self-supervised algorithm that does not require any prior information on the input image, estimation on the noise, or ground-truth data. The denoising of images~\cite{FANG2020103044} is achieved through the extraction of features from the noisy image through a convolutional neural network (CNN), and combining the edge regularisation with the total variation regularisation. The combination of CNN and low-rank representation~\cite{fu2021hyperspectral} is applied to detect anomalous pixels in hyperspectral images. The multilevel wavelet convolutional neural network~\cite{wu2020deep} is applied for restoring blurred images affected by Cauchy noise. The \emph{block matching Convolutional Neural Network} (BM-CNN)~\cite{ahn2017block} integrates a deep learning approach with the 3D block-matching method; the denoising of the stacks is predicted through a DnCNN~\cite{zhang2017beyond}, which is trained with a data set of 400 images corresponding to more than 250K training samples. A feed-forward Convolutional Neural Network smooths images, independently from the noise level, by exploiting residual learning and batch normalisation. Then, the blocks are aggregated and the image is reconstructed, as in the 3D block-matching algorithm.
 
\paragraph{Deep learning methods for image-to-image regression}
The \emph{VGG19}~\cite{simonyan2014very} introduces \emph{Convolutional Neural Networks} (CNN) pushing the depth to 16–19 weight layers, using small convolution filters of~$3 \times 3$ size, with an application to large scale images classification. The \emph{Pix2Pix}~\cite{isola2017image} method is a \emph{Generative Adversarial Network} (GAN), where the generator is a U-net~\cite{ronneberger2015u}, the discriminator is an encoding network~\cite{krizhevsky2012imagenet}, and the loss function is based on the binary cross-entropy. The \emph{deep convolutional generative adversarial network}~\cite{radford2015unsupervised} applies unsupervised learning for image classification and the generation of natural images, exploiting batch normalisation, rectified linear unit activations, and removing fully connected hidden layers.

\section{Method\label{sec:METHOD}}
We propose a novel deep learning framework for the real-time denoising of ultrasound images. Firstly, we introduce the data sets and metrics for the selection of the denoising method that best fits our requirements for ultrasound images (Sect.~\ref{SEC:COMPARISON}). Then, we optimise the parameters of the selected method to the denoising of ultrasound images (Sect.~\ref{SEC:DATASET3}). Finally, we introduce a deep learning (Sect.~\ref{SEC:REQUS}) and HPC (Sect.~\ref{SEC:HPC}) framework, which achieves real-time denoising. For a detailed discussion on the results, we refer the reader to Sect.~\ref{sec:RESULTS}.
\begin{figure*}
\centering
\begin{tabular}{cccc}
Raw image & WNNM & SAR-BM3D & \\
\includegraphics[height=0.19\textwidth]{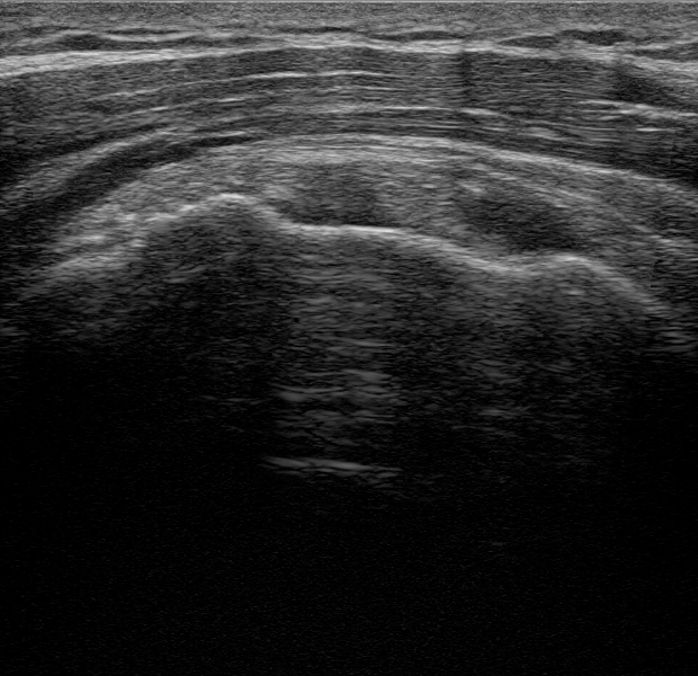} &
\includegraphics[height=0.19\textwidth]{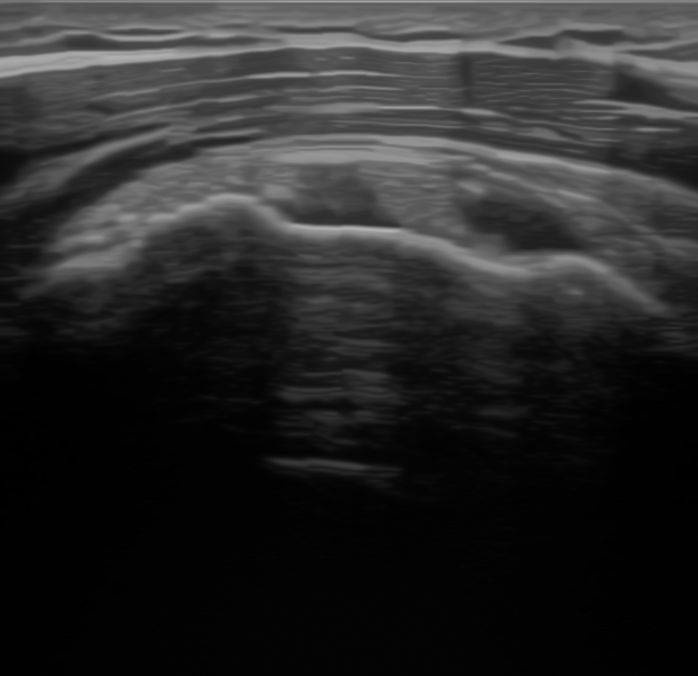} &
\includegraphics[height=0.19\textwidth]{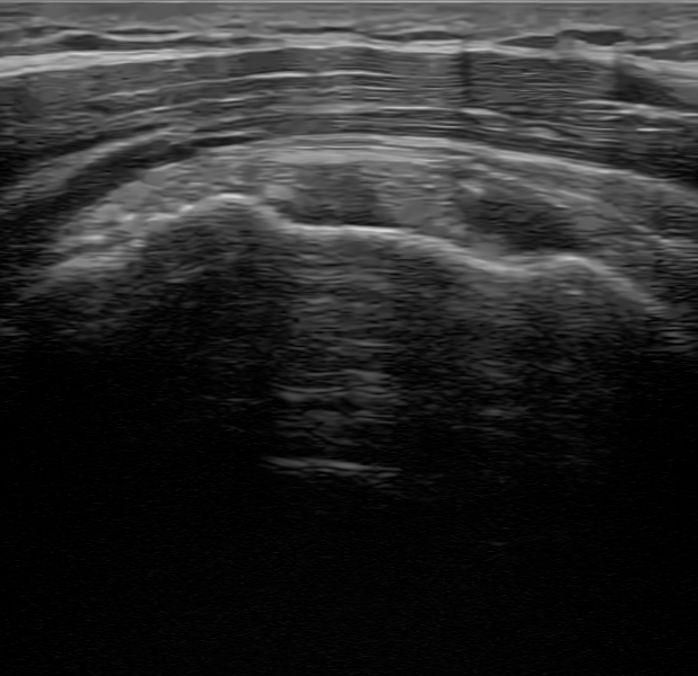} & \\
PCA-BM3D & NCSR & BM-CNN & Tuned-WNNM (Ours) \\
\includegraphics[height=0.19\textwidth]{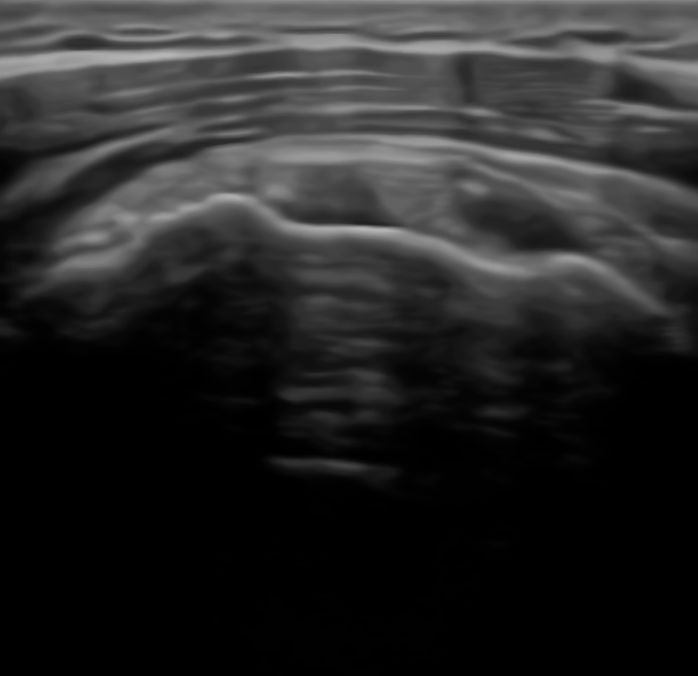} &
\includegraphics[height=0.19\textwidth]{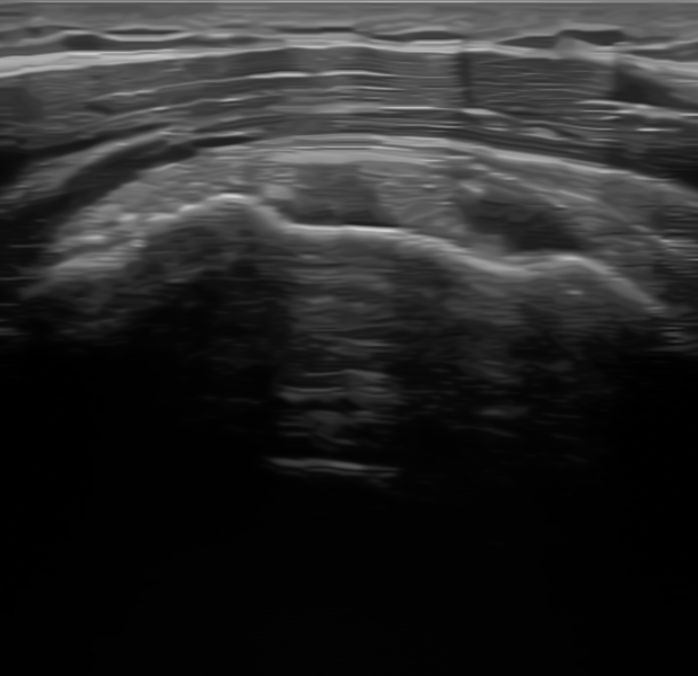} &
\includegraphics[height=0.19\textwidth]{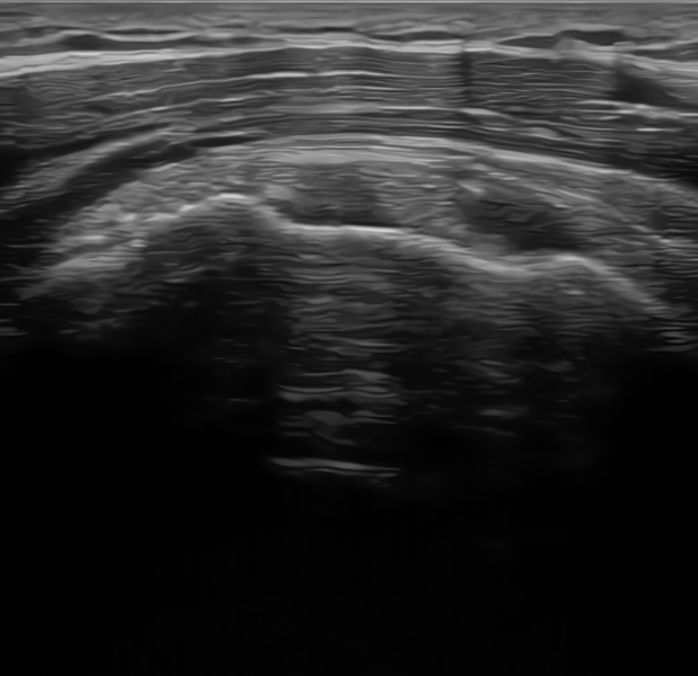} & 
\includegraphics[height=0.19\textwidth]{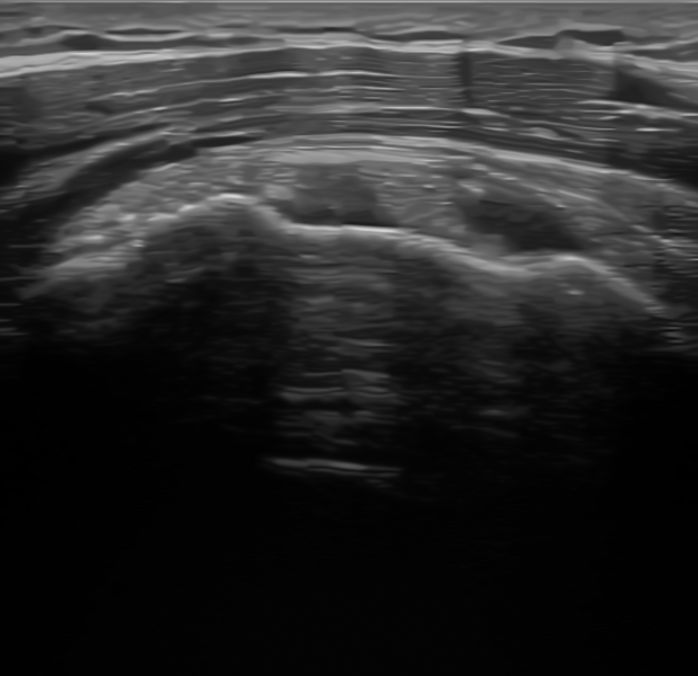}
\end{tabular}
\caption{Raw images of a muscle-skeletal district and denoised images, visualised in a scan-converted format. \label{FIG:ESA2}}
\end{figure*}
\subsection{Data sets and metrics for denoising evaluation\label{SEC:COMPARISON}}
We compare five denoising state-of-the-art methods, which are either specific for speckle noise (e.g., SAR-BM3D) or independent of the type of noise (e.g., WNNM). We consider the \emph{Esaote private data set}, which contains more than 10K ultrasound images at different resolutions, and is acquired from different (e.g., muscle-skeletal, obstetric, abdominal) anatomical districts. On this data set, we compare the performance of different denoising methods applied to ultrasound images, and analyse the performance of the proposed framework, through the training and the prediction of the learning-based network, with a specialisation to anatomic districts.
 
As \emph{quantitative metrics}, we consider the \emph{peak-signal-to-noise ratio} (PSNR) and the \emph{structural similarity index measure} (SSIM) for the comparison of the raw input with the target denoised image provided by the proposed framework. Furthermore, we integrate the quantitative metrics with a qualitative discussion on the quality of the denoised images, in terms of blurring and edge preservation.
\begin{table}[t]
\caption{Execution time computed as an average value of a set of Esaote images at~$ 600 \times 485$ resolution.\label{TAB:computcost}}
\centering
\resizebox{\columnwidth}{!}{
\begin{tabular}{c|ccccc}
\textbf{Method} & WNNM & SAR-BM3D & BM-CNN & NCSR & PCA-BM3D \\ \hline
\begin{tabular}{c}
\textbf{Execution} \\
\textbf{time [s]}
\end{tabular}
& 215 &~$\mathbf{55}$ & 356 & 380 & 95
\end{tabular}
}
\end{table}
\subsection{Tuned-WNNM\label{SEC:DATASET3}}
According to the results in Sect.~\ref{sec:RESULTS}, WNNM has been selected as the best denoising method among five state-of-the-art methods belonging to the spectral, low-rank, and deep learning classes. To improve the quality of the denoised image, we propose a novel approach to the tuning of WNNM parameters, and we refer to this method as \emph{tuned-WNNM}. 

Given a pixel~$x$, the \emph{patch}~$P_{x}$ is the set of pixels in the neighbourhood of~$x$; each pixel of the image has a related patch. The \emph{search window} is the set of patches selected for searching the closest ones to a reference patch, under a certain metric. The \emph{stack}, or \emph{3D block}, is the set of patches that are similar to a reference patch; these patches are stored in a 3D structure, and the redundancy of the stack is exploited to remove the noise. Within this setting, the tuning of these parameters (i.e., search window, stack, patch size) improves the results of tuned-WNNM with respect to WNNM. Our framework maximises the performance of the denoising method; in fact, the real-time requirement is achieved by the network prediction, while the denoising is applied offline for the generation of the training data set.

\subsection{Real-time denoising with deep learning\label{SEC:REQUS}}
The main requirements of a denoising algorithm for ultrasound images are the magnitude of the removed noise, edge preservation, and real-time computation. The tuned-WNNM satisfies these requirements, except for the execution time, which does not satisfy the real-time need of ultrasound applications. To achieve a real-time computation and to maintain the good results of the tuned-WNNM method in terms of denoising and edges preservation, we identify two strategies. We develop (i) a computationally optimised version of the tuned-WNNM method, exploiting HPC tools, CPUs and GPUs, and low-level programming languages. We design and implement (ii) a deep learning framework that uses the tuned-WNNM as an instance of denoising methods.
 
The implementation of a computationally optimised, and potentially real-time, version of the tuned-WNNM is a very tough requirement; the main iterative cycle of the algorithm is not parallelizable, due to the dependency of the data among the iterations. Furthermore, the cubic computational cost for the evaluation of the SVD of each block is no further optimisable. The real-time requirement needs a specific hardware-based implementation, and any modification to the method requires a new implementation of the parallel algorithm. This approach needs dedicated and more expensive hardware, which is in contrast with the cheapness of the ultrasound acquisition.
\begin{figure*}
\centering
\begin{tabular}{cccc}
Raw image & WNNM & SAR-BM3D & \\
\includegraphics[height=0.15\textwidth]{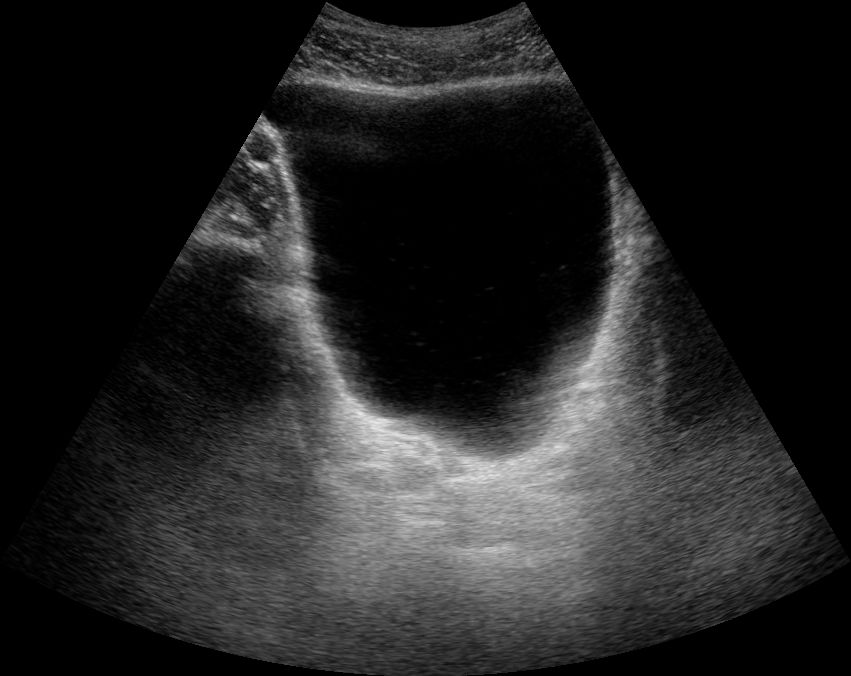} &
\includegraphics[height=0.15\textwidth]{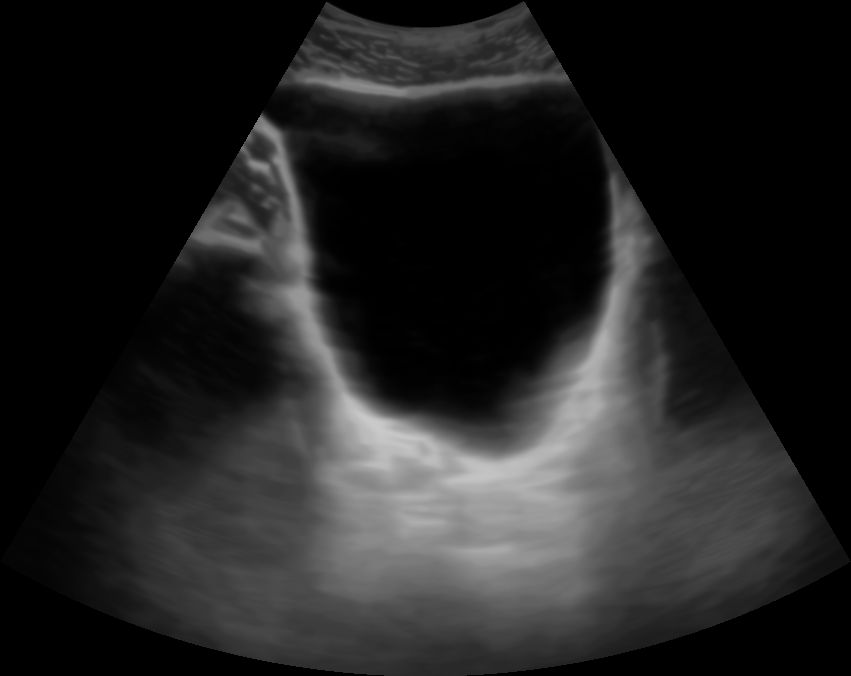} &
\includegraphics[height=0.15\textwidth]{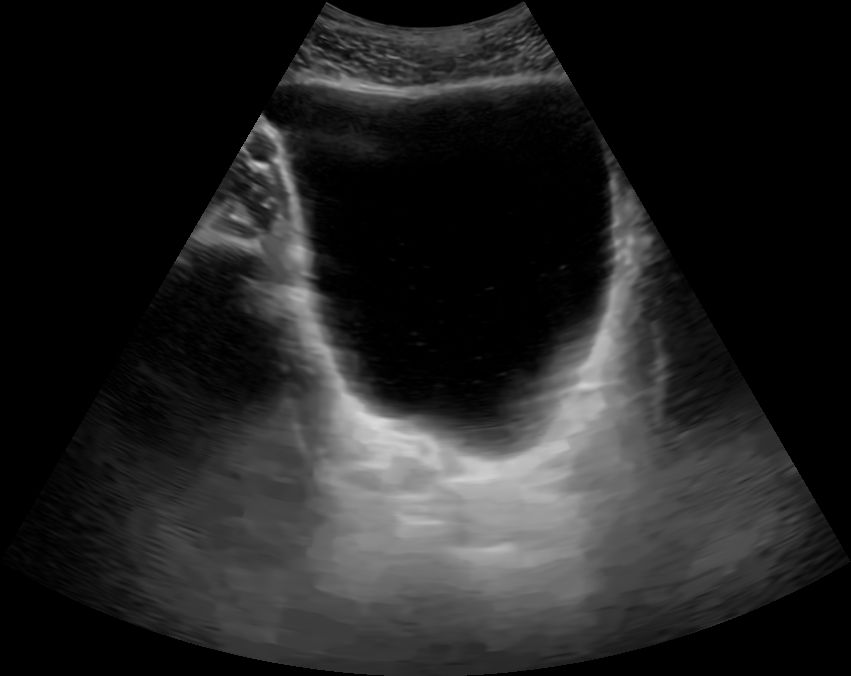} \\
PCA-BM3D & NCSR & BM-CNN & Tuned-WNNM (Ours) \\
\includegraphics[height=0.15\textwidth]{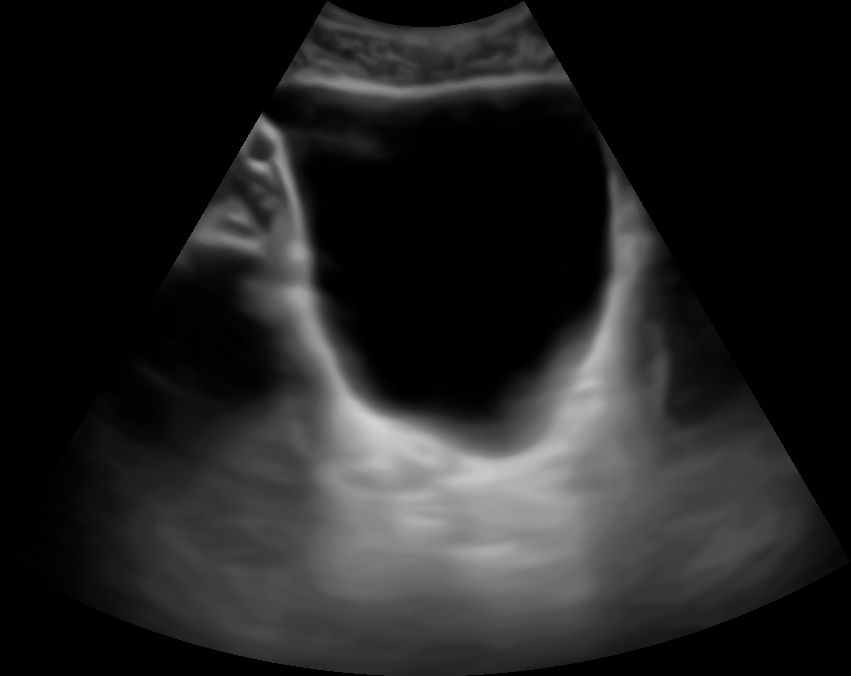} &
\includegraphics[height=0.15\textwidth]{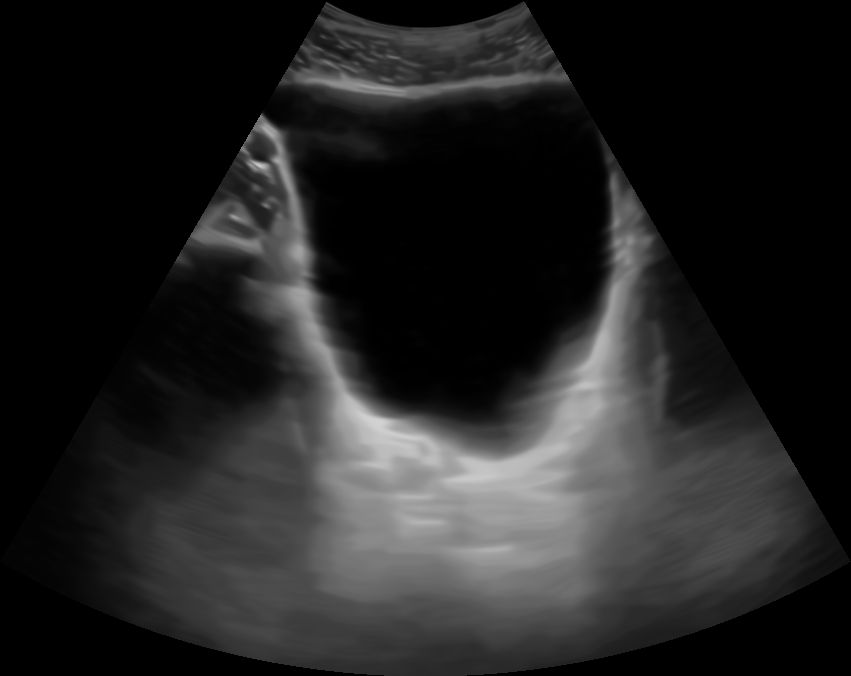} &
\includegraphics[height=0.15\textwidth]{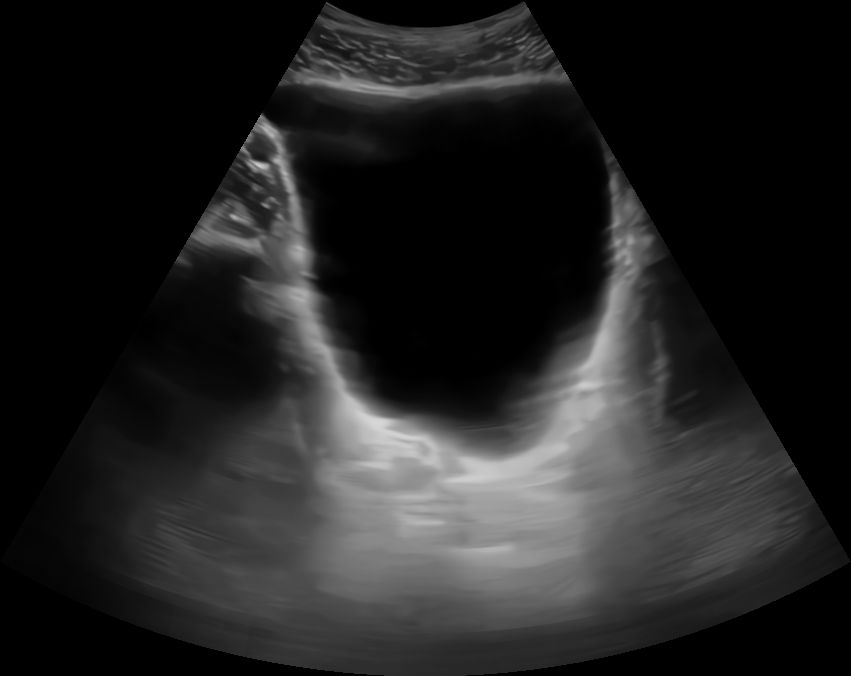} &
\includegraphics[height=0.15\textwidth]{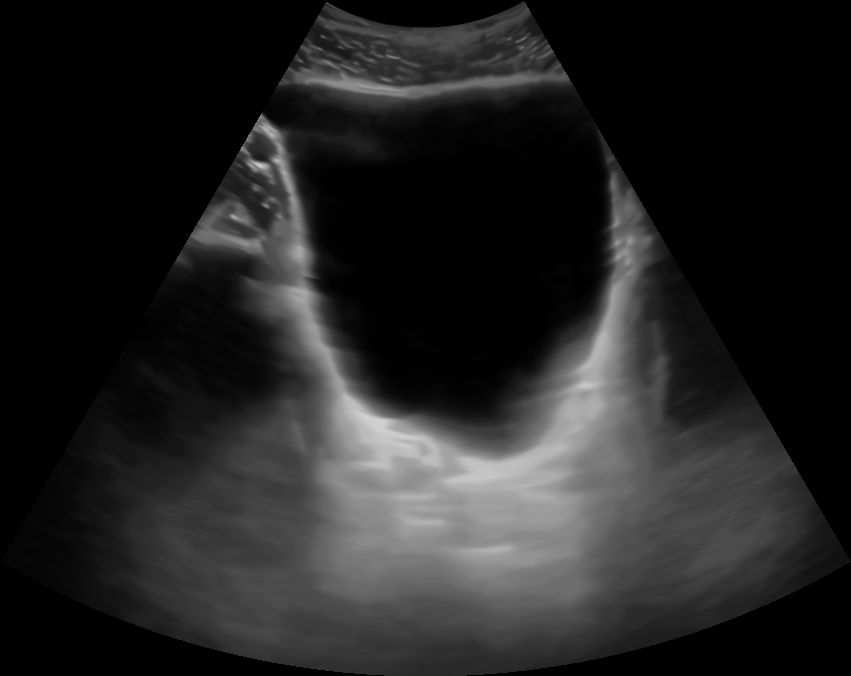}
\end{tabular}
\caption{Raw data set of an abdominal district and denoised images, visualised in a scan-converted format. \label{FIG:ESA}}
\end{figure*}
\paragraph{Proposed approach and contributions}
The proposed real-time denoising is based on the training of a neural network to learn and replicate the tuned-WNNM behaviour. In the first phase, the network is trained on a data set of ultrasound images of the same district. The data set for the training of the learning method is composed of a set of couples of ultrasound images: the input (i.e., the raw image) and the target (i.e., the image denoised with the tuned-WNNM filter). The ground truth is not available in ultrasound applications; for this reason, the target of the learning method is the output of the tuned-WNNM filter. Then, the trained network provides the denoised output through a real-time prediction of the test images. 
As the main contribution, the proposed deep-learning framework is general in terms of the input data, i.e., type of noise (e.g., speckle, Gaussian noise), resolution (e.g., isotropic, anisotropic) of the input images, acquisition methodology, and the anatomical district. Our deep learning framework is also general in terms of building blocks and parameters: since different methods (e.g., NCSR, SAR-BM3D, custom methods) have good performances, the generality of our framework allows us to use a different denoising algorithm and to exploit its different characteristics in terms of denoising and edge preservation. Alternative denoising methods can be used for different types of noise (e.g., speckle, Gaussian noise) and signals (e.g., 3D images, time-dependent ultrasound videos).

In our approach, we specialise the training phase to specific anatomical districts or types of noise. For instance, we train a specific network for each district, thus obtaining a more precise result when predicting the denoised image, as each network is specialised to the input anatomical features. We also improve the WNNM denoising in terms of real-time computation based on offline training. The real-time computation depends only on the execution time of the network prediction. Furthermore, the offline training can be improved with new data, a-priori and/or additional information on the data (e.g., input anatomical district, noisy type/intensity, image resolution, acquisition methodology/protocol). The update of the existing training data set is always performed offline, through the addition of new images after expert validation of the denoising results.
\begin{figure*}
\centering
\begin{tabular}{cccc}
Raw image & WNNM & SAR-BM3D & \\
\includegraphics[height=0.13\textwidth]{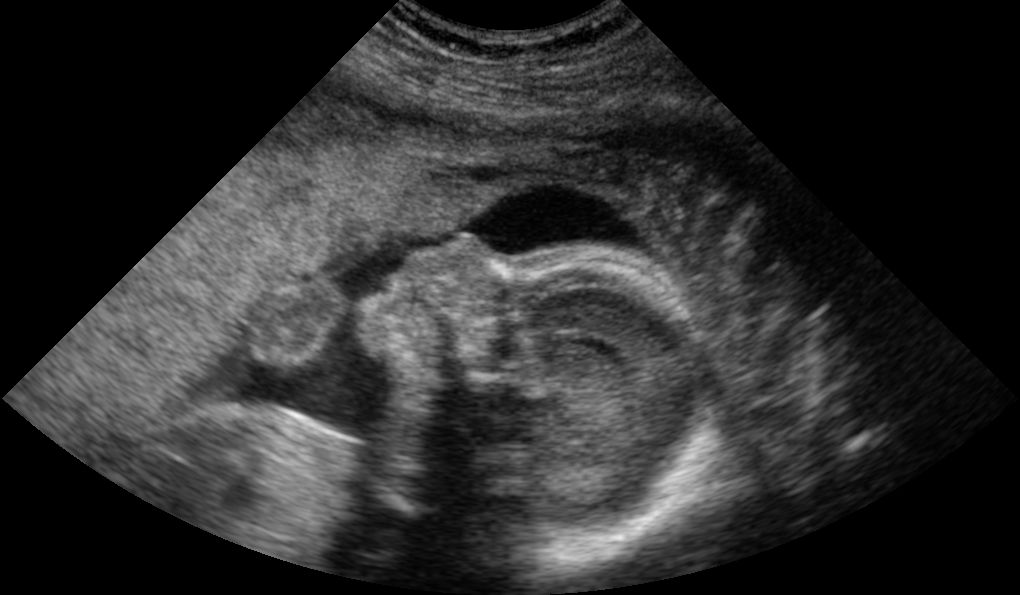} &
\includegraphics[height=0.13\textwidth]{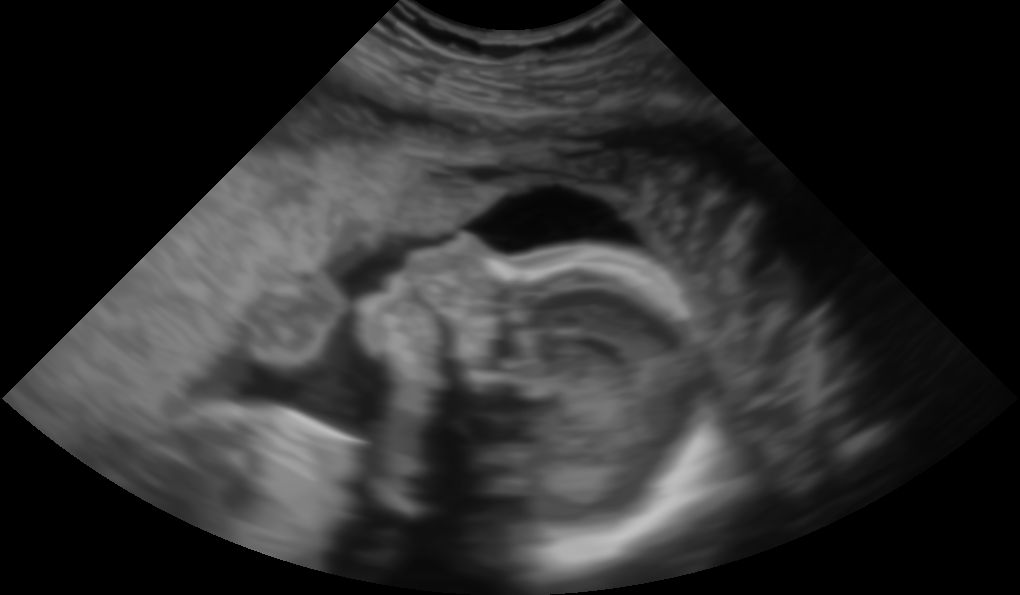} &
\includegraphics[height=0.13\textwidth]{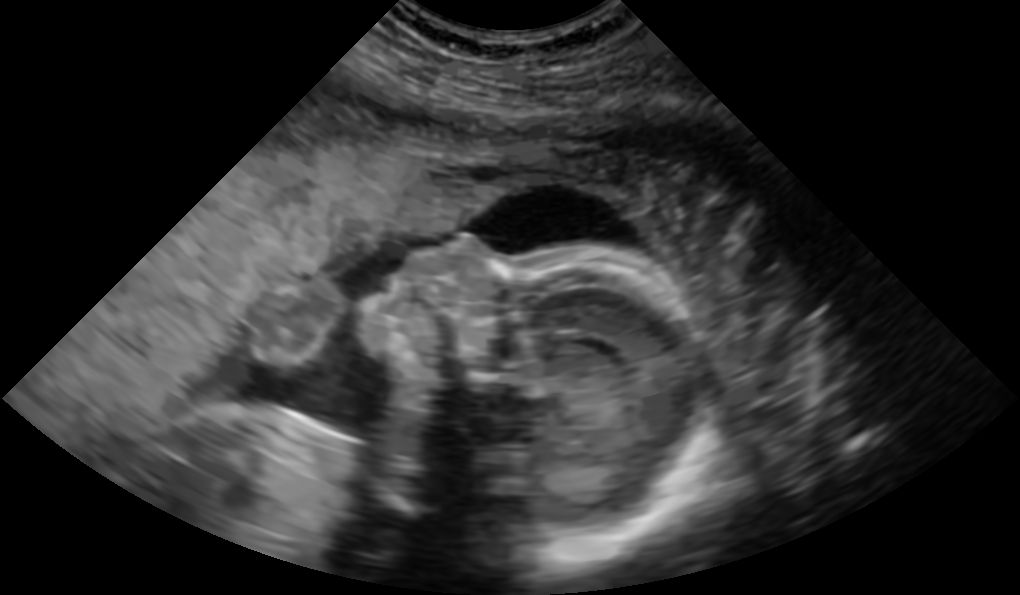} \\
PCA-BM3D & NCSR & BM-CNN & Tuned-WNNM (Ours) \\
\includegraphics[height=0.13\textwidth]{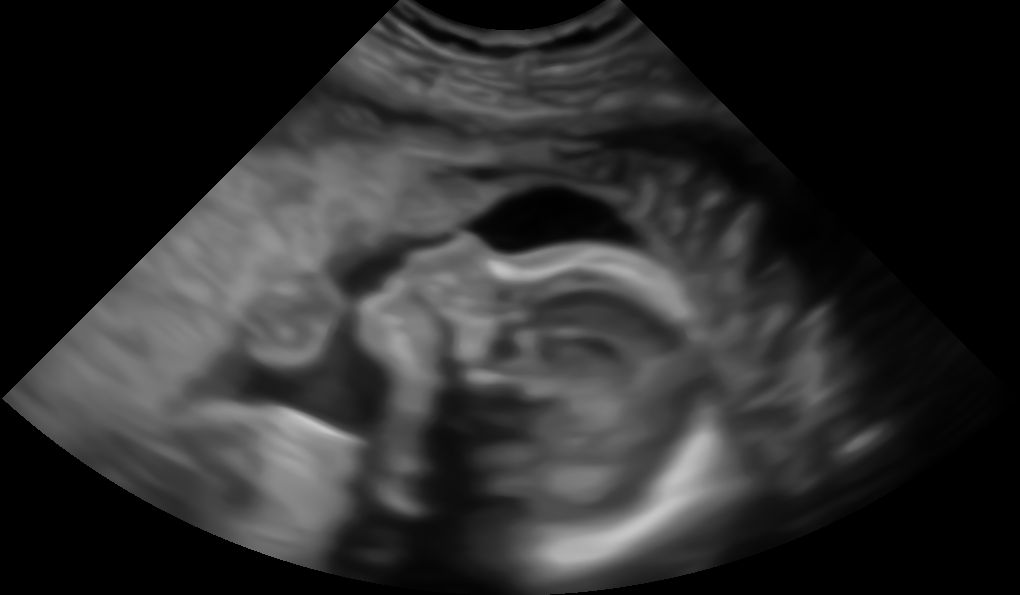} &
\includegraphics[height=0.13\textwidth]{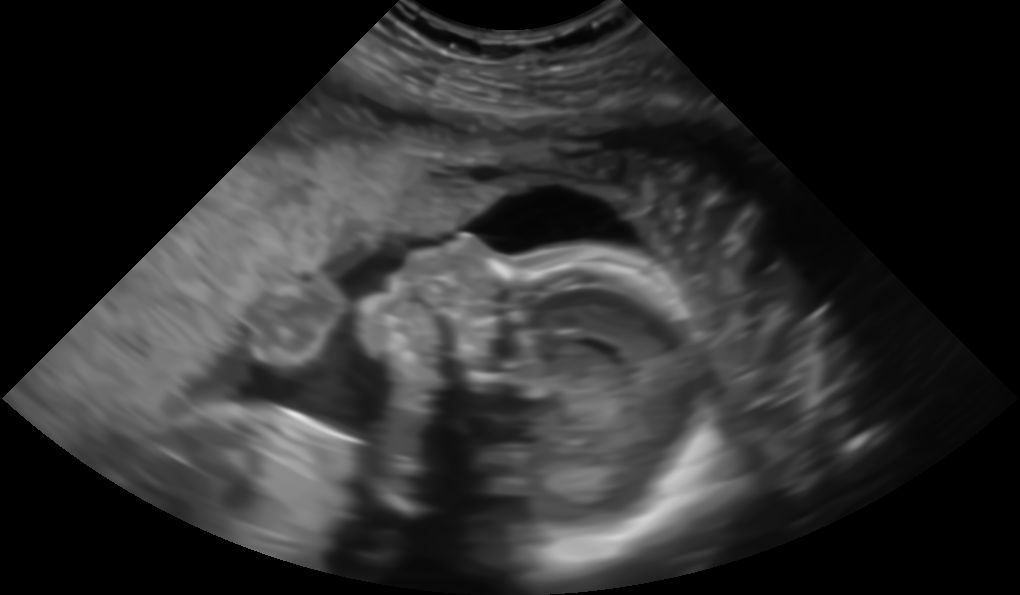} &
\includegraphics[height=0.13\textwidth]{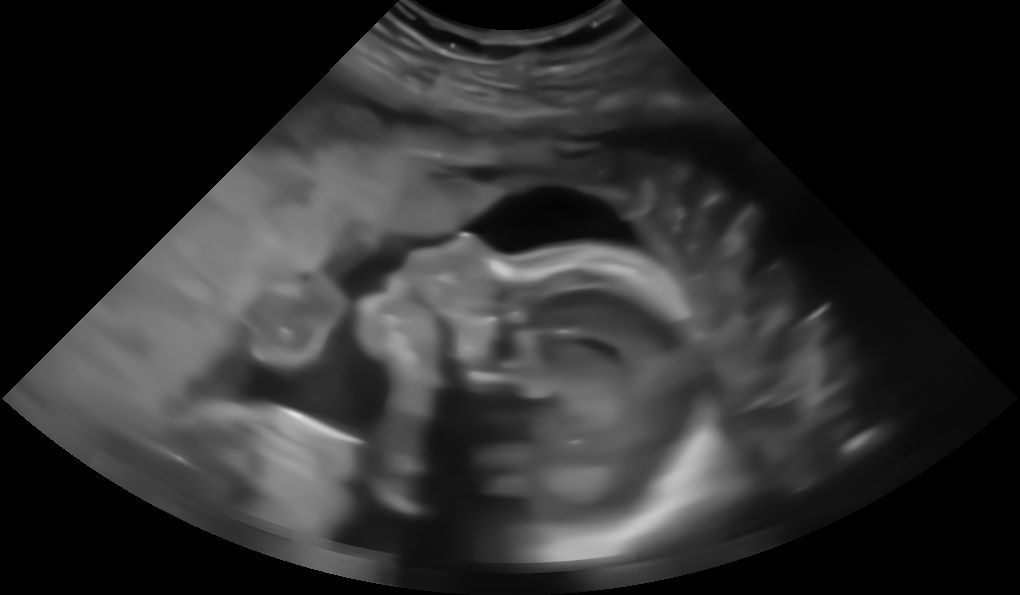} &
\includegraphics[height=0.13\textwidth]{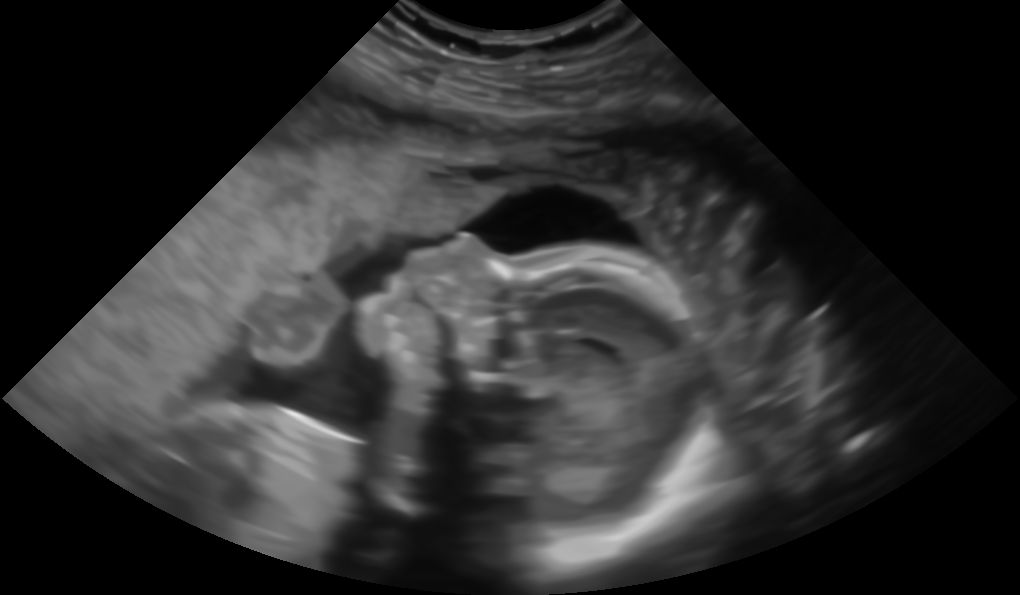}
\end{tabular}
\caption{Raw data set of an obstetric district and denoised images, visualised in a scan-converted format. \label{FIG:ESA3}}
\end{figure*}
\paragraph{Deep learning architecture\label{sec:DEEP-LEARNING-ARCHITECTURE}}
To evaluate the proposed framework, we analyse several networks and perform an image-to-image regression; among them (Sect.~\ref{SEC:RELWORK}), we select Pix2Pix, which guarantees good results in terms of learning. We specialise Pix2Pix to ultrasound images, with two updates: (i) the introduction of a validation data set of the same district of the training data set, which forces the exit condition when the validation error increases, and (ii) the introduction of padding and masking pre-processing operations, which allow us to deal with images of different resolution, without an image resize that would imply a distortion artefact. A comparison between Pix2Pix and \emph{Matlab} CNN is discussed in Sect.~\ref{SEC:NNcomparison}.

\paragraph{Training data sets\label{SEC:DATASET4}}
We generate and test different data sets, by varying the number of images for the training phase, and the anatomical district for the prediction phase. The custom Pix2Pix network is trained on four data sets of obstetric images, with respectively: (a) 500, (b) 1500, (c) 3500, and (d) 5000 images. Each data set is composed of the input images (i.e., the raw Esaote data set) and the target images (i.e., the corresponding images denoised with the tuned-WNNM). The validation data set is composed of an additional set of different images (i.e., about 10\% of the training data set) of the same district. Then, we evaluate each of the four networks (i.e., the networks trained with a different number of images) with two different test data sets of 50 images each, respectively from the (i) obstetric and (ii) muscular anatomical districts. For each test data set, we compute the quantitative PSNR and SSIM metrics between the prediction of the network and the expected target; furthermore, the experts visually evaluate the prediction results.

\subsection{HPC framework for learning\label{SEC:HPC}}
We define an HPC implementation of the proposed deep learning framework, taking advantage of a large ultrasound data set with 5K ultrasound images, and of the CINECA-Marconi100 cluster, exploiting both CPUs (IBM POWER9 AC922) and GPUs (NVIDIA Volta V100). Given a training data set, composed of raw images and the corresponding denoised images, we implement a parallel and distributed deep learning framework in TensorFlow2. Then, we define a batch file for the execution of the deep learning framework on the cluster, which specifies the number of nodes, CPUs, GPUs, and memory of the cluster. Through the proposed HPC framework, we train multiple networks with large data sets in a reasonable time for the target medical application, thus increasing the specialisation to anatomical districts, and consequently the accuracy of the deep learning framework. The HPC framework generates a network model that is stored and used for predicting the output results. Furthermore, we can improve the offline training with new data, a-priori and/or additional information on the input data (e.g., input anatomical district, noisy type/intensity, image resolution, acquisition methodology/protocol). The training data set can be periodically updated with the denoised images after the expert validation of the denoising results.

\section{Results\label{sec:RESULTS}}
We present the results of denoising methods with a specialisation to ultrasound images (Sect.~\ref{SEC:QUANTITATIVE3}), a comparison between baseline and tuned-WNNM (Sect.~\ref{SEC:SPECIALISATIONRESULTS}), deep learning (Sect.~\ref{SEC:QUALITATIVE4}) and HPC (Sect.~\ref{SEC:EXECTIME}) framework. Finally, we compare the real-time denoising with different neural network architectures (Sect.~\ref{SEC:NNcomparison}).
\begin{figure*}
\centering
\begin{tabular}{cccc}
Input image & Tuned-WNNM: (a) &Tuned-WNNM: (b) & Tuned-WNNM: (c) \\
\includegraphics[width=0.23\textwidth]{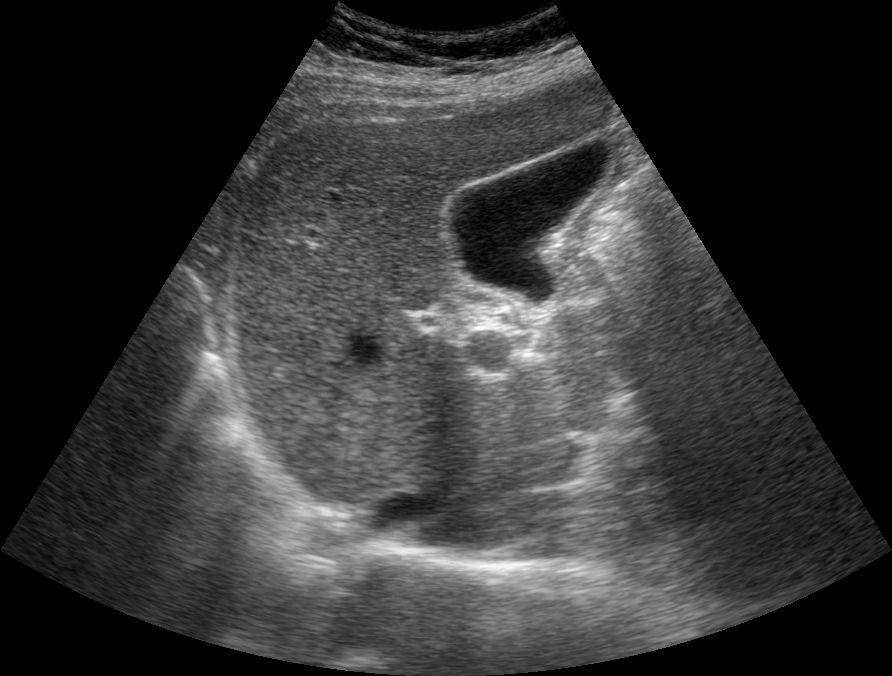} &
\includegraphics[width=0.23\textwidth]{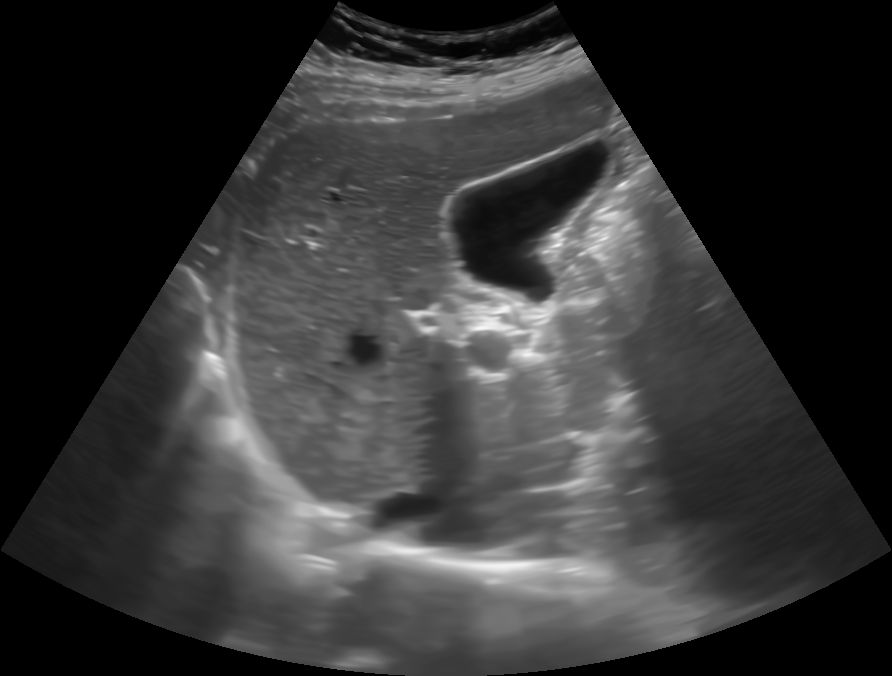} &
\includegraphics[width=0.23\textwidth]{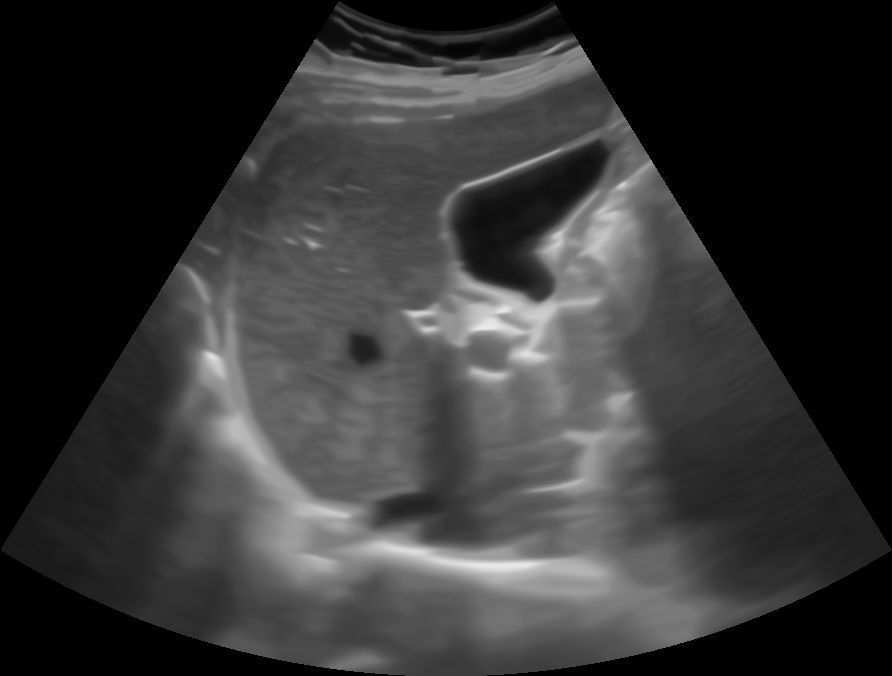} &
\includegraphics[width=0.23\textwidth]{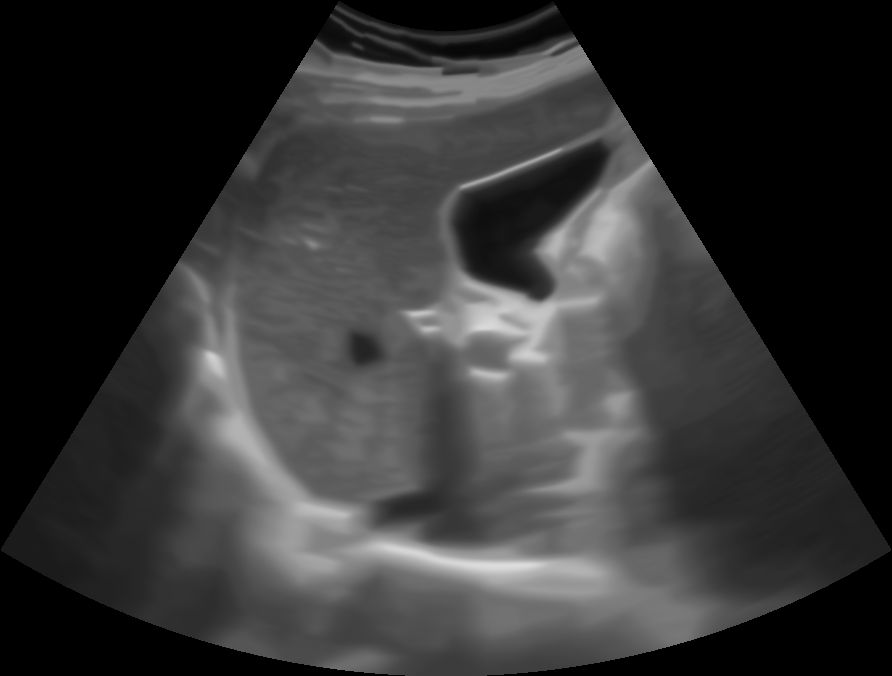}
\end{tabular}
\caption{Ultrasound image of an abdominal district and denoised images achieved by applying the tuned-WNNM and varying the denoising intensity from low (a) to high (c).\label{FIG:PARAMWNNM}}
\end{figure*}
\subsection{Comparison of denoising methods\label{SEC:QUANTITATIVE3}}
We evaluate the denoising results on different anatomical districts of the Esaote data set (Figs.,~\ref{FIG:ESA2},~\ref{FIG:ESA},~\ref{FIG:ESA3}): WNNM, NCSR, and PCA-BM3D have been judged as the best methods in terms of denoising, and WNNM outperforms all the other methods in terms of edge preservation and enhancement. In particular, WNNM well preserves the edges of the muscular fibres (Fig.~\ref{FIG:ESA2}) and the internal organs (Figs.~\ref{FIG:ESA},~\ref{FIG:ESA3}). The output of SAR-BM3D shows a granular effect, which negatively affects the preservation of the anatomical features, and BM-CNN generates artefacts, which are typical of a deep learning approach. According to these results, we select WNNM as the best method for the denoising of ultrasound images. However, we underline that the other methods have their characteristics in terms of denoising and edge preservation, and they could be included in the framework as alternative denoising algorithms.
\paragraph*{Execution time\label{SEC:TIME3}}
Our tests (Table~\ref{TAB:computcost}) are executed with Matlab R2020a, on a workstation with 2 Intel i9-9900KF CPUs (3.60GHz), 32 GB RAM, and none of these methods achieves real-time computation. In particular, WNNM takes more than three minutes to process a~$600 \times 485$ image, and the fastest method (i.e., SAR-BM3D) takes about one minute; however, real-time computation in an ultrasound environment requires a processing time in the order of a few milliseconds. This result motivates the proposed development of a deep learning framework for the real-time denoising of ultrasound images, further optimised with a HPC framework (Sect.~\ref{SEC:HPC}).

\subsection{Tuned-WNNM for US images\label{SEC:SPECIALISATIONRESULTS}}
We implement the tuned-WNNM through the optimisation of the following parameters.
The number of patches is no more limited by the step value (e.g., 1 patch every 2 or 3 pixels) and we assign a patch for each pixel; this parameter allows us to increase the number of processed patches, thus improving the data redundancy. The block-matching algorithm is now performed every iteration, instead of one every two iterations; the selection of the searching window and the size of the stack are now larger than previous work. These parameters allow us to improve the measure of the similarity among 3D blocks and the accuracy of the denoising method.

Furthermore, we specialise the tuned-WNNM method to ultrasound images, by varying the denoising intensity through a parameter that affects the threshold of the singular values of the SVD. Increasing this parameter, the method improves in terms of removed noise, though introducing a low blurring effect. To select the best tuning for denoising intensity, we select the output image that best fits the medical requirements, among three different levels of denoising intensity (Fig.~\ref{FIG:PARAMWNNM}). In particular, Fig.~\ref{FIG:PARAMWNNM}(b) shows the best result as a compromise between noise removal, edges preservation, and blurring effect; in fact, it preserves the geometry of the internal tissues, while enhancing the edges of the anatomical structures. 

\subsection{Learning-based denoising}\label{SEC:QUALITATIVE4}
\begin{figure*}
\centering
\begin{tabular}{c|cc}
Raw image & Prediction:(a) & Prediction: (b) \\
\includegraphics[width=0.25\textwidth]{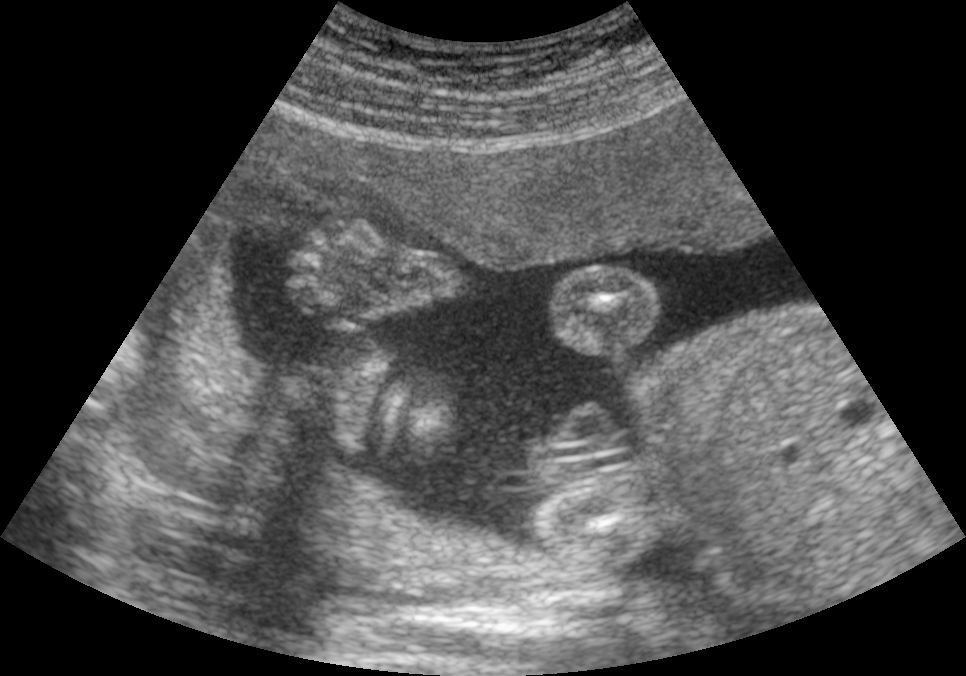} &
\includegraphics[width=0.25\textwidth]{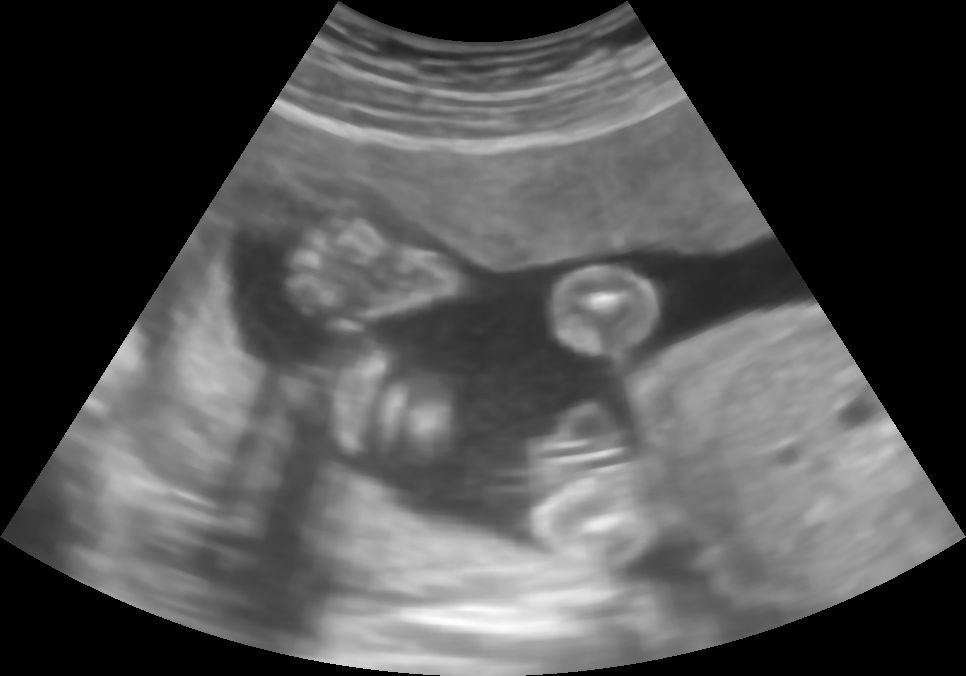} &
\includegraphics[width=0.25\textwidth]{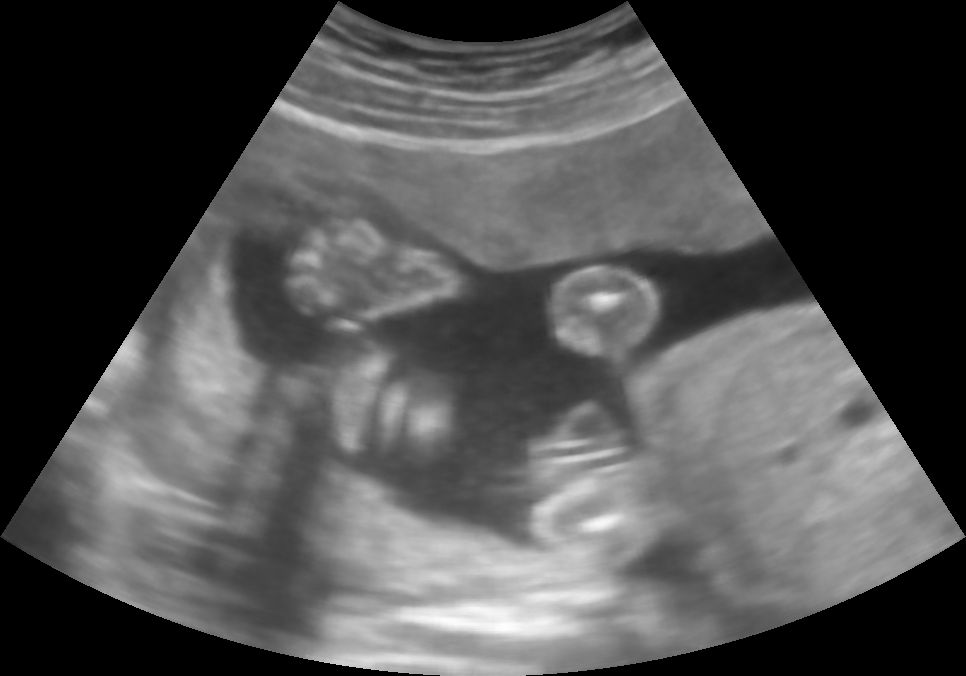} \\
Target & Prediction: (c) & Prediction: (d) \\
\includegraphics[width=0.25\textwidth]{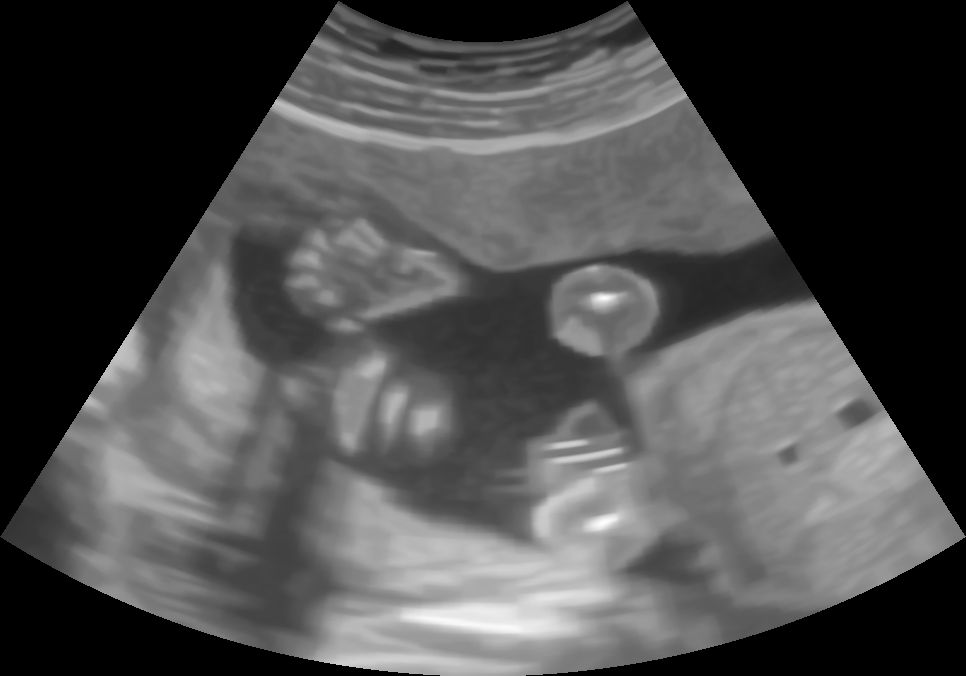} &
\includegraphics[width=0.25\textwidth]{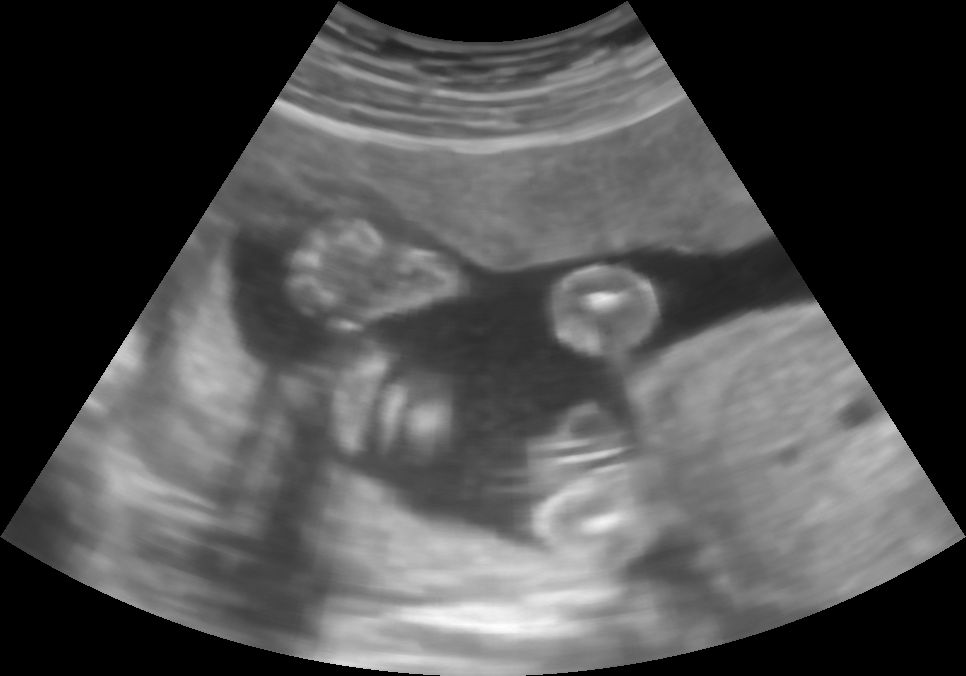} &
\includegraphics[width=0.25\textwidth]{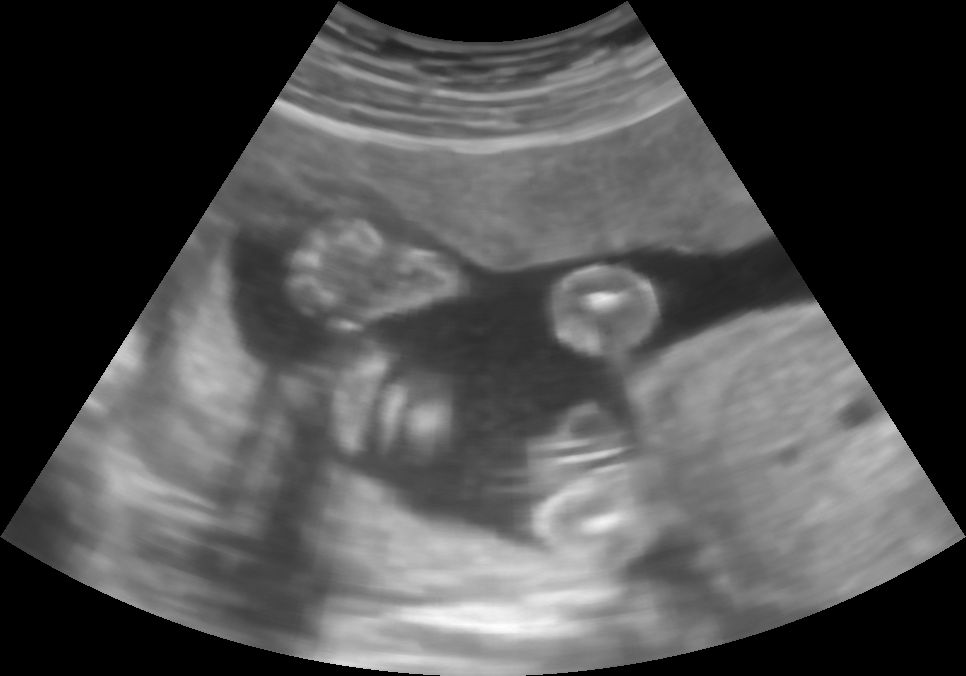}
\end{tabular}
\caption{Raw, target, and prediction images, related to the obstetric data set (i). Training set: (a) 500 images, (b) 1500 images, (c) 3500 images, (d) 5000 images (Sect.~\ref{SEC:DATASET4}).\label{FIG:GANOB}}
\end{figure*}
\begin{figure*}
\centering
\begin{tabular}{c|cc}
Raw image & Prediction:(a) & Prediction: (b) \\
\includegraphics[width=0.20\textwidth]{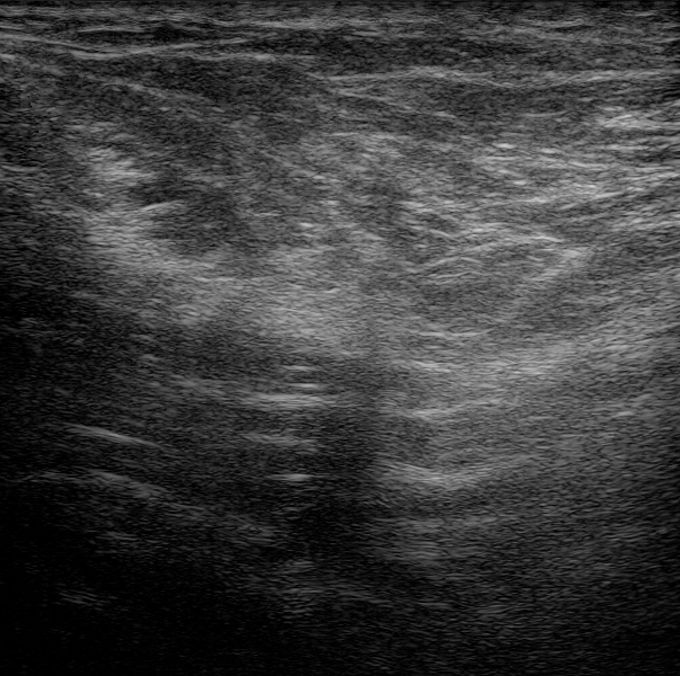} &
\includegraphics[width=0.20\textwidth]{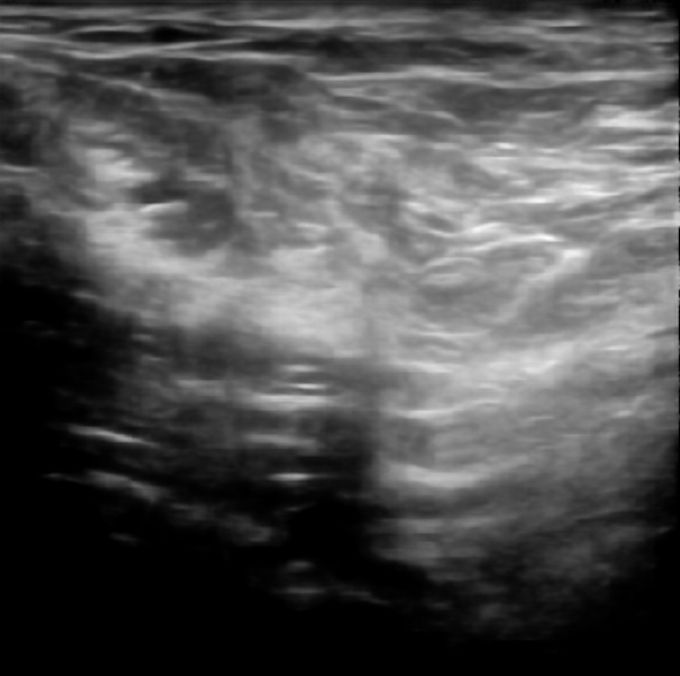} &
\includegraphics[width=0.20\textwidth]{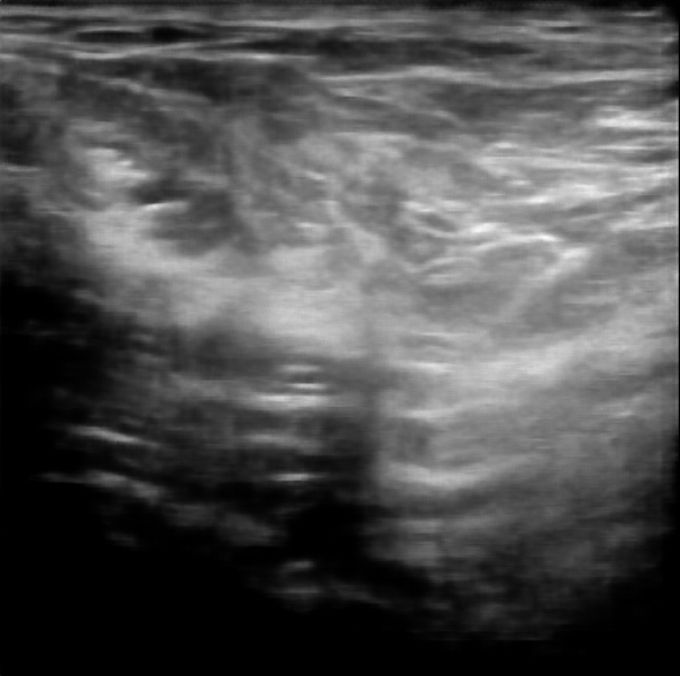} \\
Target & Prediction: (c) & Prediction: (d) \\
\includegraphics[width=0.20\textwidth]{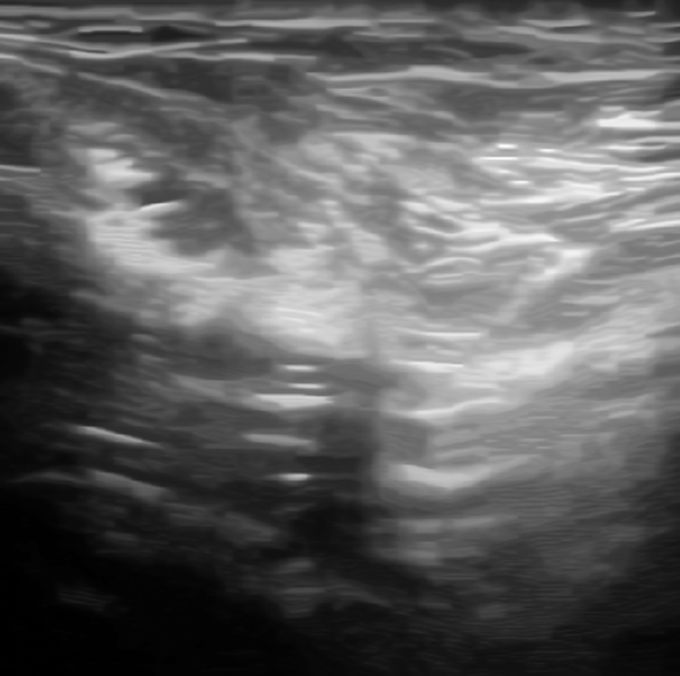} &
\includegraphics[width=0.20\textwidth]{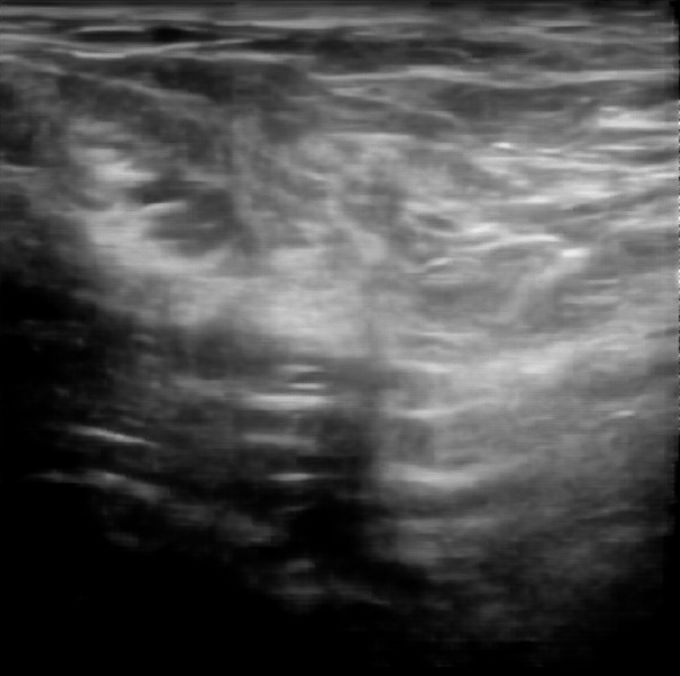} &
\includegraphics[width=0.20\textwidth]{predictionSaveMSK_2_3444.jpg}
\end{tabular}
\caption{
Raw, target, and prediction images, related to the muscle-skeletal data set (ii). Training set (Sect.~\ref{SEC:DATASET4}) with (a) 500, (b) 1500, (c) 3500, and (d) 5000 images.\label{FIG:GANMSK}}
\end{figure*}
\begin{figure}
\centering
\begin{tabular}{cc}
\includegraphics[height=52pt]{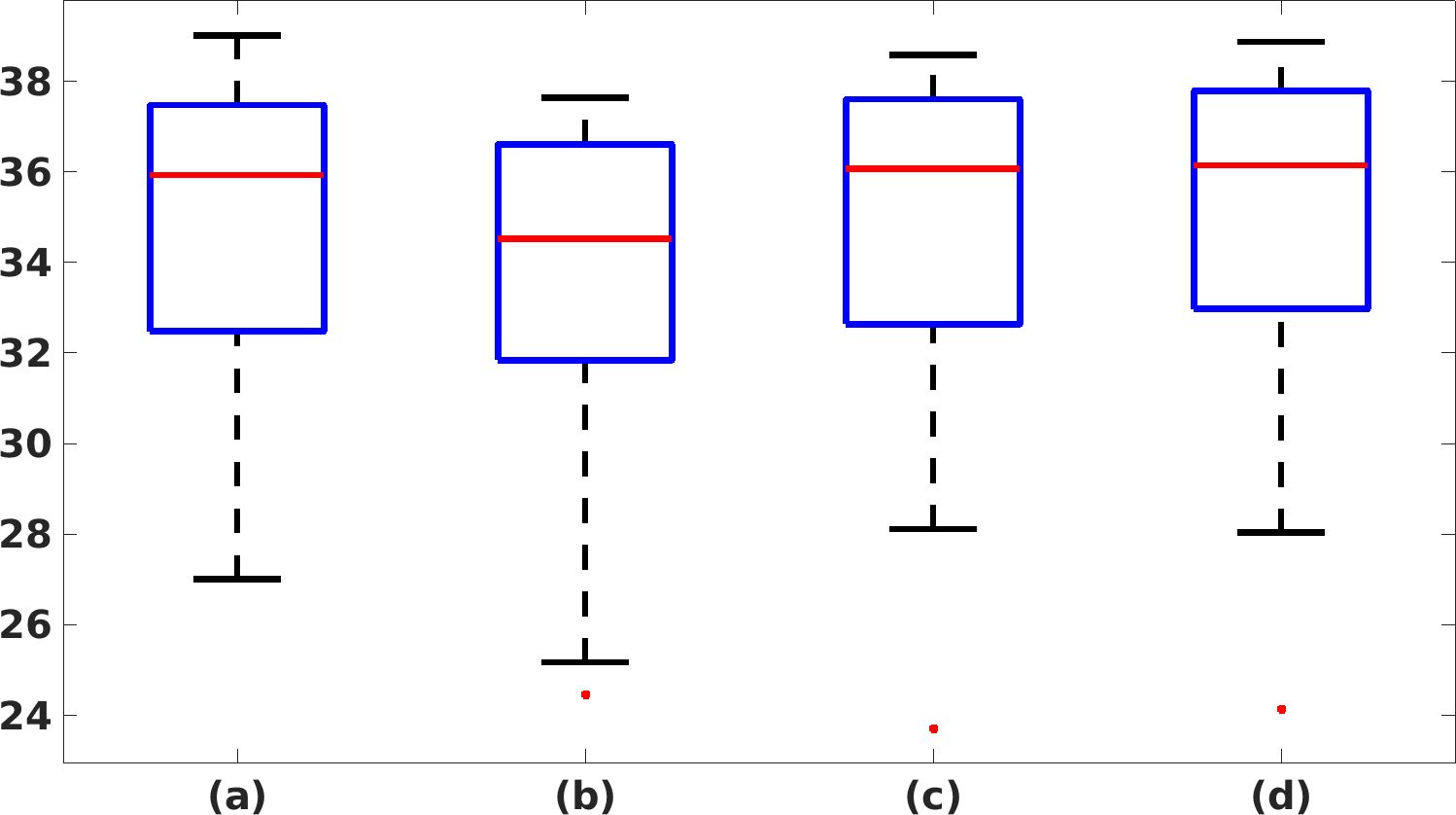} &
\includegraphics[height=52pt]{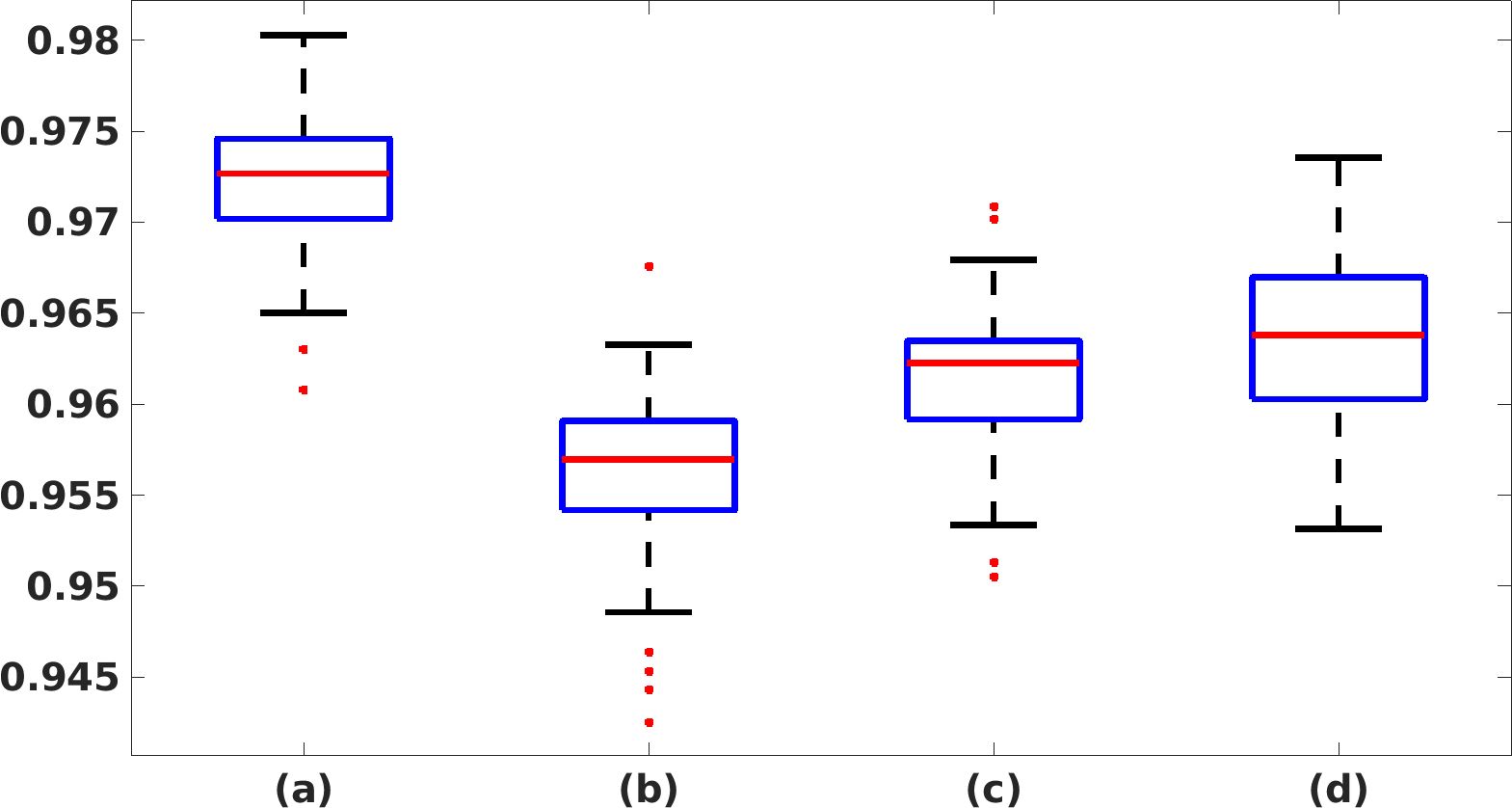} \\
\includegraphics[height=52pt]{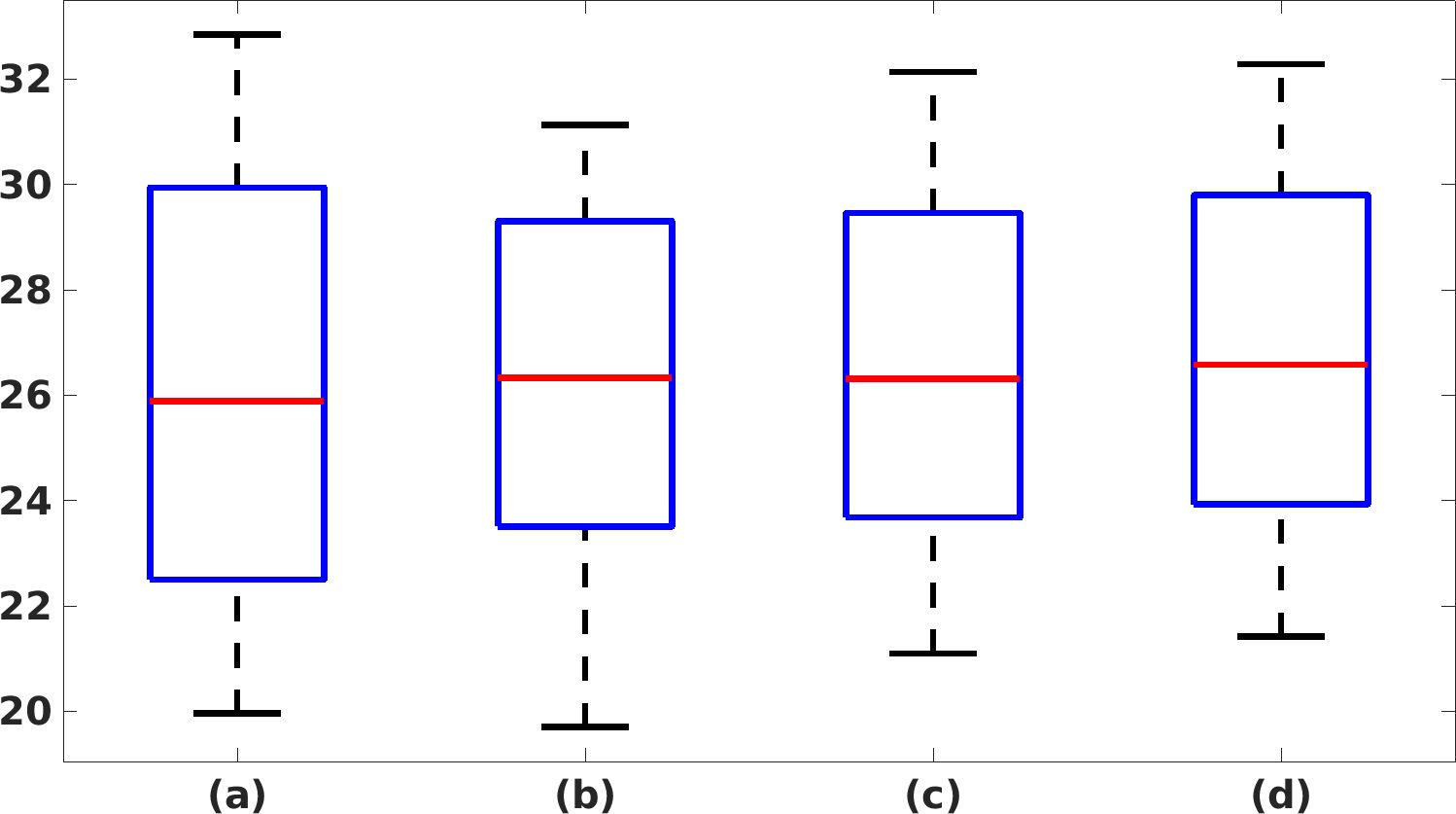} &
\includegraphics[height=52pt]{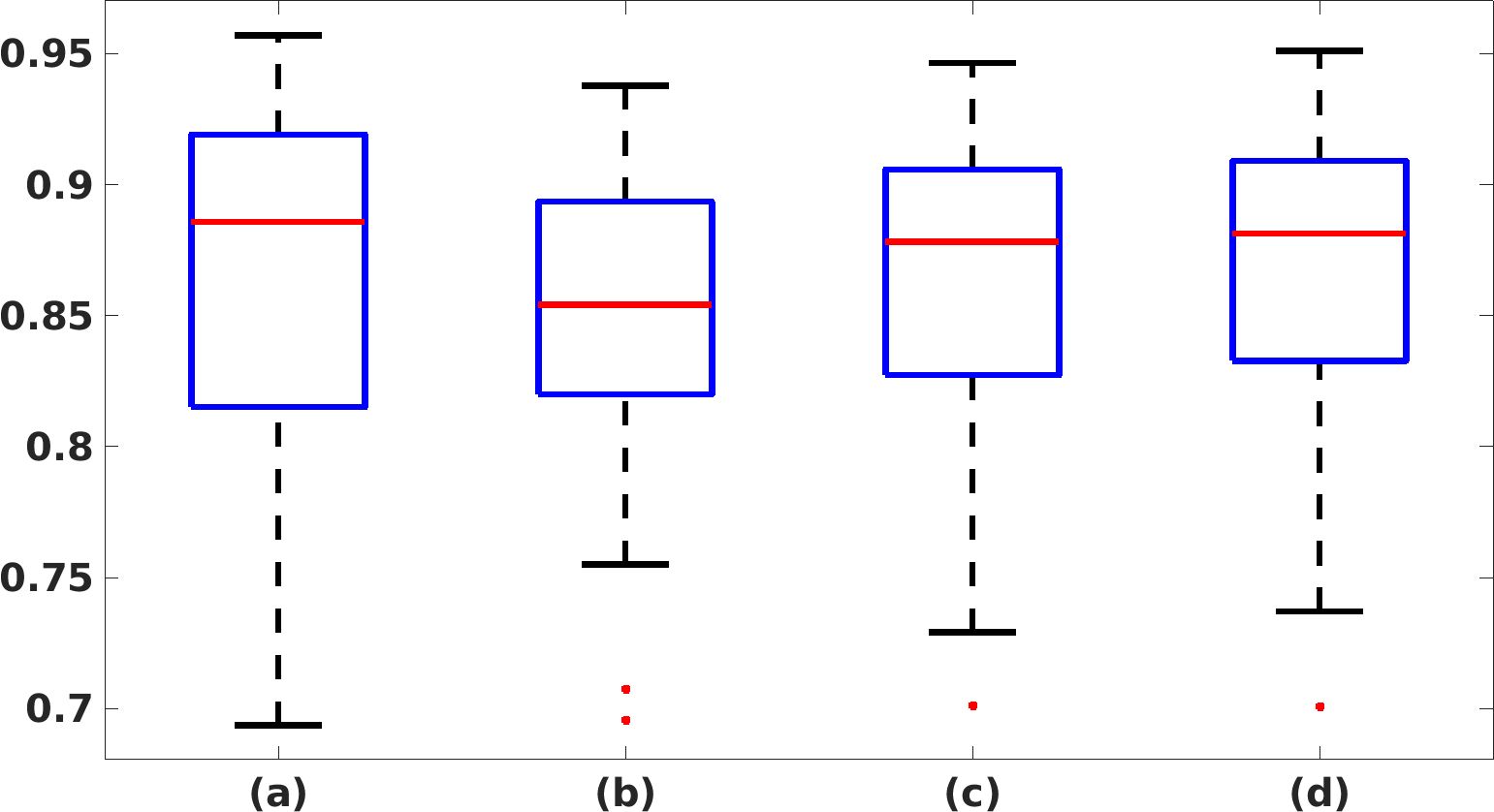}
\end{tabular}
\caption{PSNR and SSIM boxplot for each of the four training data set (i.e., (a-d)) and the two test data sets (Sect.~\ref{SEC:DATASET4}): (top-left) PSNR, obstetric test data set; (top-right) SSIM, obstetric test data set; (bottom-left) PSNR, muscle-skeletal test data set; (bottom-right) SSIM, muscle-skeletal test data set.\label{FIG:GANMETRICS}}
\end{figure}
\begin{figure*}
\centering
\begin{tabular}{ccc}
Input image & Target image & \\ 
\includegraphics[width=0.20\textwidth]{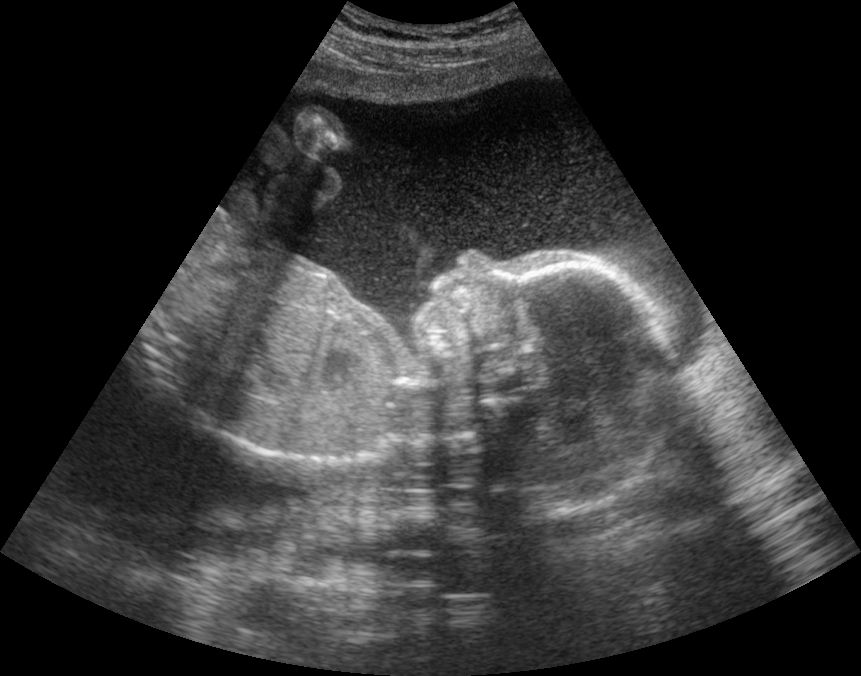} &
\includegraphics[width=0.20\textwidth]{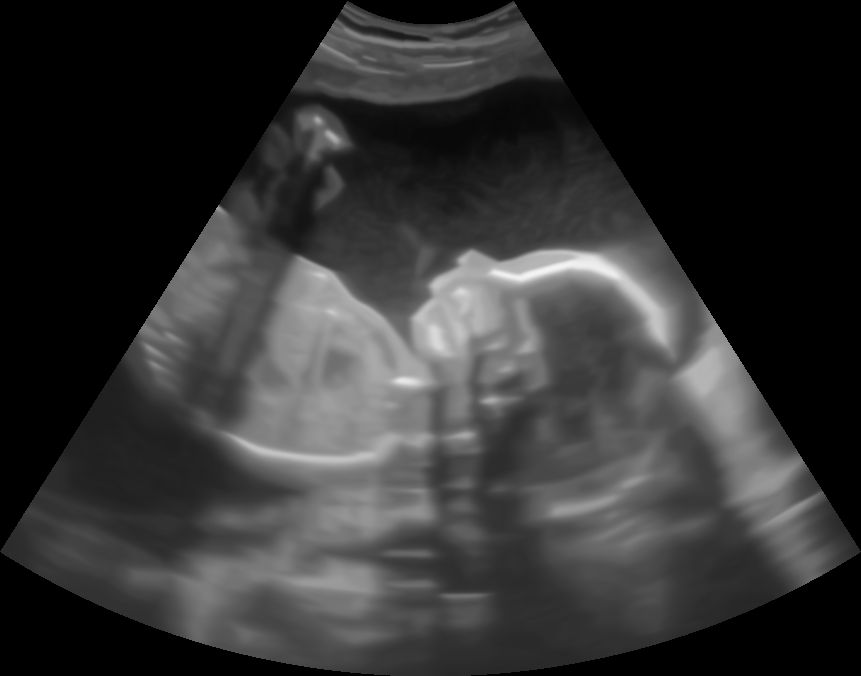} & \\
Single district (a) & Single district (b) & Multiple district \\
\includegraphics[width=0.20\textwidth]{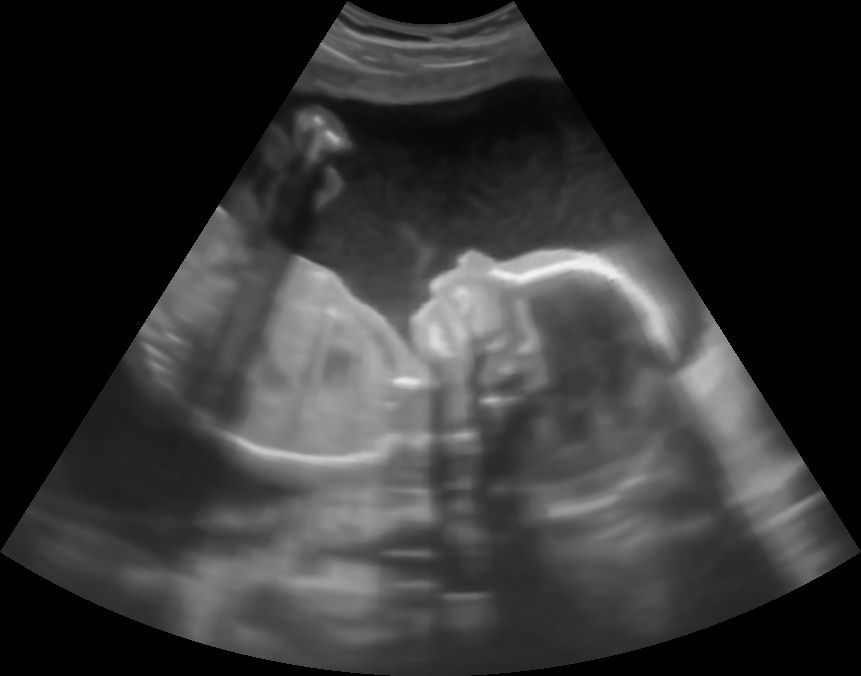} &
\includegraphics[width=0.20\textwidth]{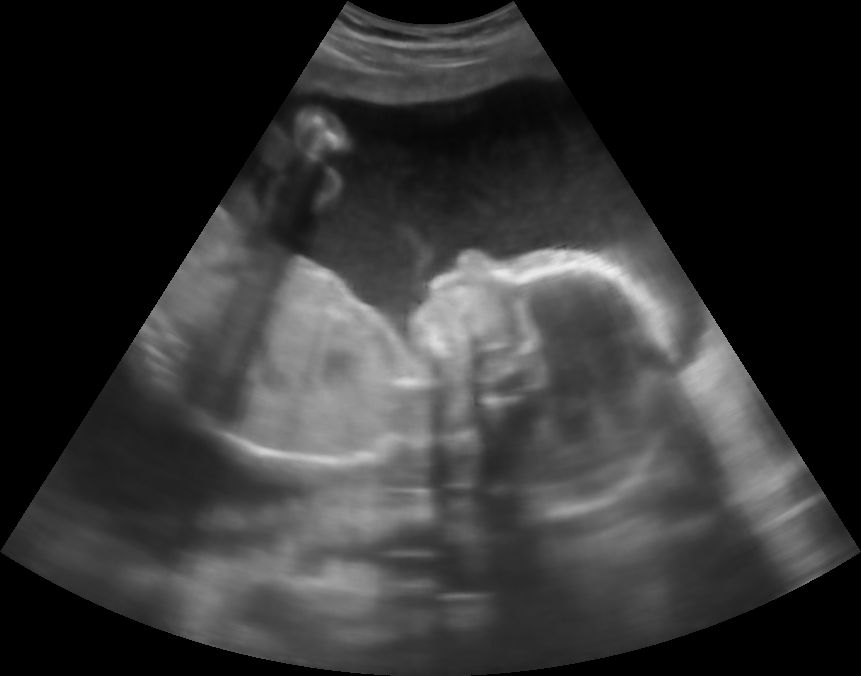} &
\includegraphics[width=0.20\textwidth]{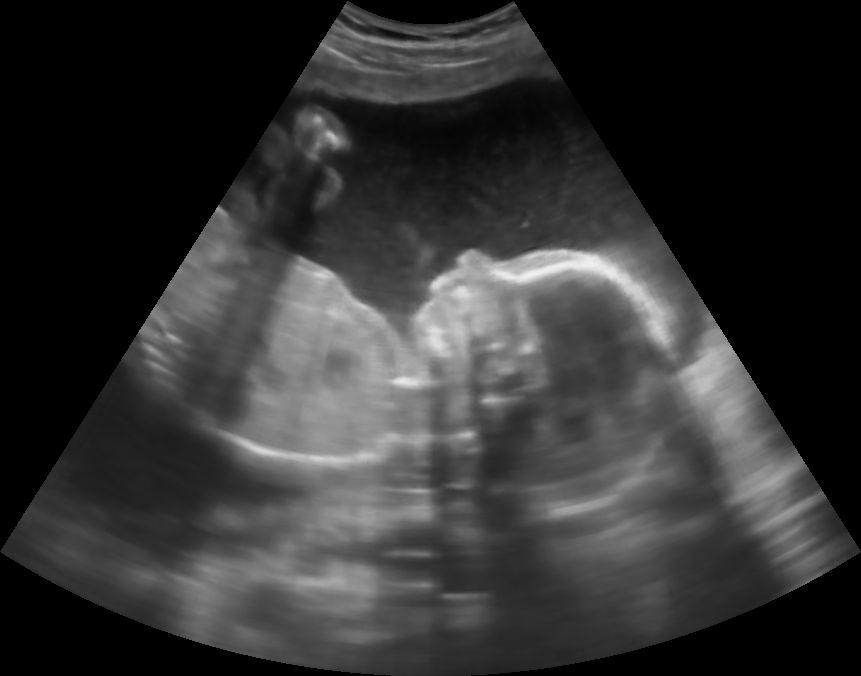}
\end{tabular}
\caption{Prediction results of the obstetric district, with the networks trained with 500 (a) and 1500 (b) images from the obstetric district, and 1500 images from multiple districts (500 obstetric, 500 cardiac, 500 muscle-skeletal images).\label{FIG:multidis1}}
\end{figure*}

\paragraph{Qualitative results}
Regarding the deep learning framework, and the large ultrasound data set (Sect.~\ref{SEC:DATASET4}), Fig.~\ref{FIG:GANOB} shows the prediction results of the four networks, when tested with obstetric images (i). The predicted images are very close to the target image in all four cases; the edges and the grey-scale values are well reproduced by the network. Furthermore, the predictions do not generate artefacts or spurious patterns. Varying the number of images of the training data set from 500 to 5K (Figs.~\ref{FIG:GANOB}(a-d)), the predicted images are slightly better than the target denoised images. Nevertheless, the results are good even with a small training data set of 500 images. Fig.~\ref{FIG:GANMSK} shows the prediction results of the four networks when tested with muscle-skeletal images (ii). Predicting the output images with the networks trained with obstetric images (Figs.~\ref{FIG:GANMSK}(a-d)), the results are slightly worse than the corresponding case in Fig.~\ref{FIG:GANOB}, even if the predicted images do not show any artefact of pattern repetition. These networks are trained with images from a different (i.e., obstetric) district, with different anatomical features. This result confirms that each district requires a specific network and that a single network for all the districts gives lower quality results.
\paragraph{Quantitative results}
Table~\ref{TAB:PSNRTEST} reports the quantitative metrics (Sect.~\ref{SEC:DATASET3}) computed between the target and the predicted images. The network trained with 5K images (d) tested with obstetric images (i) has a median PSNR and SSIM value of 36.13 and 0.964, respectively, while the same network tested with muscle-skeletal images (ii) has a median PSNR and SSIM value of 26.58 and 0.881. Both the metrics have a very slight improvement when passing from a training set of 500 to a training set of 3500 images, confirming the results of the qualitative analysis. An additional increase of the size of the training data set to 5K images further improves the quantitative results for both the test data sets.
Fig.~\ref{FIG:GANMETRICS} shows the box plot of the PSNR and SSIM metrics for four training data sets and two test data sets. Increasing the number of images of the training data set, the range of the metrics tends to decrease; this behaviour has a lower variability on the prediction of the output image.
These results confirm that a network specialised in a single anatomical district reaches the best denoising quality. The prediction of muscle-skeletal images from a network trained with obstetric images highly reduces the performance of our framework; in fact, the network learns that replicates not only the denoising algorithm itself but also its adaptation to the anatomic structures and features of each district.
\begin{table}[t]
\centering
\caption{With reference to the four training data sets and the two test data sets (i.e., obstetric: \emph{Ob.}, and muscle-skeletal: \emph{Msk.}) described in Sect.~\ref{SEC:DATASET4}, we report the PSNR and SSIM metrics computed between the target and the prediction images, as median value among the 50 test images. \label{TAB:PSNRTEST}}
\begin{tabular}{l|cc|cc}
\begin{tabular}[c]{@{}c@{}}\textbf{Metrics}\end{tabular} & \multicolumn{2}{c|}{PSNR} & \multicolumn{2}{c}{SSIM} \\ \hline
\begin{tabular}[c]{@{}c@{}}\textbf{Test data set}\end{tabular} & Ob. & Msk. & Ob. & Msk. \\ \hline
\textbf{Training with}			& 		&		&			&\\
(a) 500 images    & 35.93 & 25.88 & 0.973 & 0.886 \\
(b) 1500 images    & 34.52 & 26.33 & 0.957 & 0.854 \\
(c) 3500 images    & 36.07 & 26.31 & 0.962 & 0.878 \\
(d) 5000 images    & 36.13 & 26.58 & 0.964 & 0.881
\end{tabular}
\end{table}
\paragraph{Single versus multiple districts}
We compare the prediction results of our framework with three different training data sets. The first two data sets have 500 and 1500 ultrasound images of the same district (i.e., the obstetric one), respectively. The third one is composed of 1500 images of different districts; in particular, we select 500 images from the cardiac, obstetric, and muscle-skeletal districts. Due to the different resolutions of the images, the padding has been applied to obtain the same input resolution for each network. We evaluate the prediction results on four test data sets: the first three are composed of 50 images from the cardiac, obstetric, and muscle-skeletal districts, respectively. The fourth is composed of 50 images randomly selected from the aforementioned three districts. The prediction results (Fig.~\ref{FIG:multidis1} and Table~\ref{TAB:multidis}) show that the networks trained with obstetric images (i.e., the single district networks) give the best results with the obstetric test data set: the predicted image of the single district network shows fewer scattering artefacts than the multiple districts network. Also, the single district networks have better results in terms of quantitative metrics: adding further images from different districts to the training data set worsens the results; in fact, the single district network with 500 obstetric images has a PSNR value of 35.93, while the multiple districts network has a PSNR value of 33.70.

\begin{figure*}
\centering
\begin{tabular}{ccc}
Input image & Target image & \\
\includegraphics[width=0.20\textwidth]{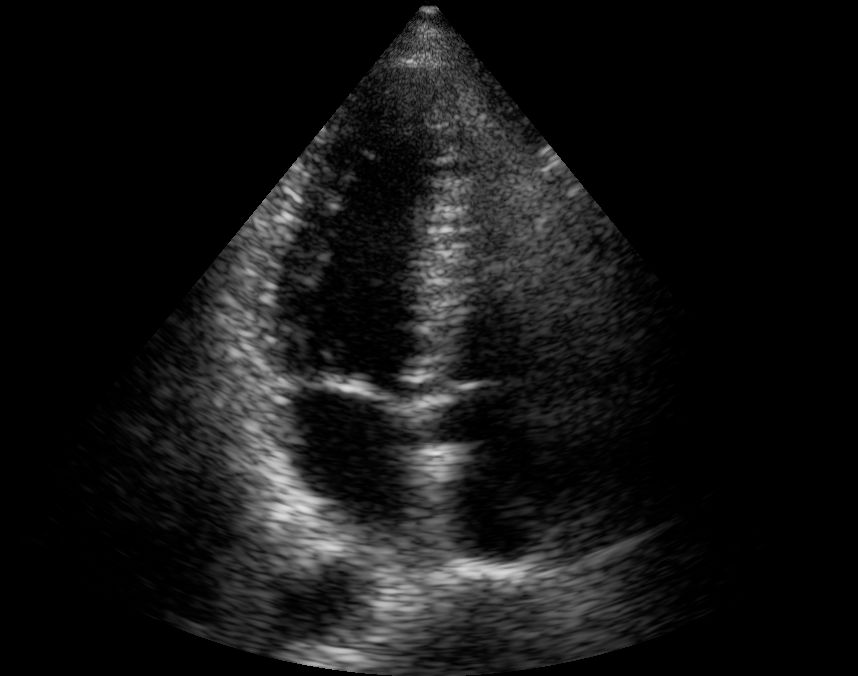} &
\includegraphics[width=0.20\textwidth]{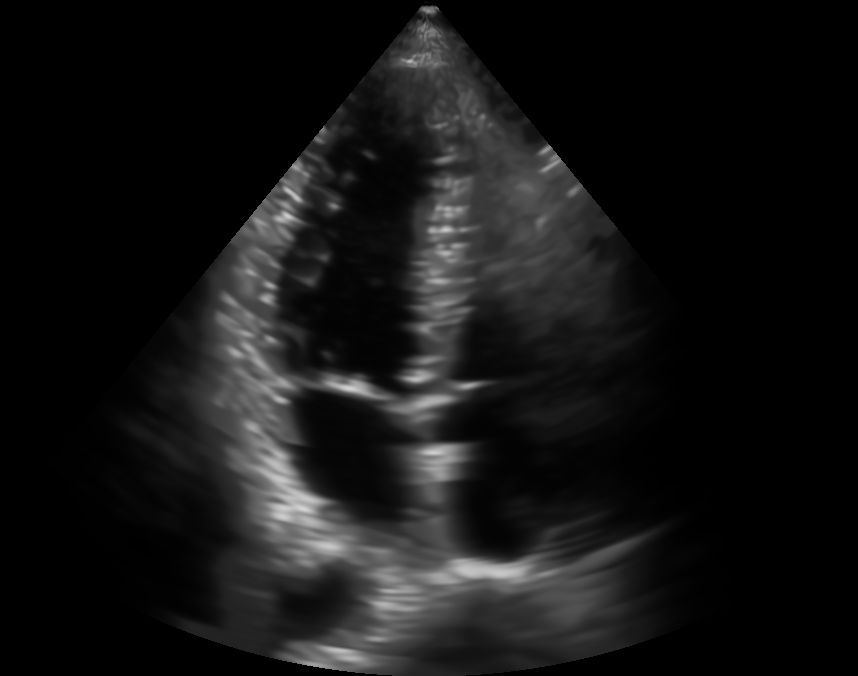} & \\
Single district (a) & Single district (b) & Multiple district \\
\includegraphics[width=0.20\textwidth]{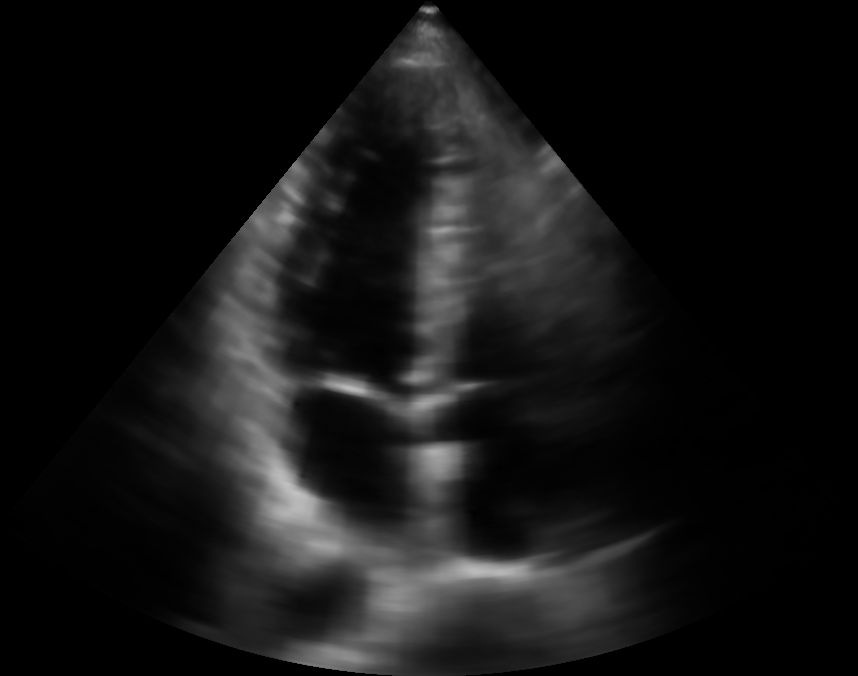} &
\includegraphics[width=0.20\textwidth]{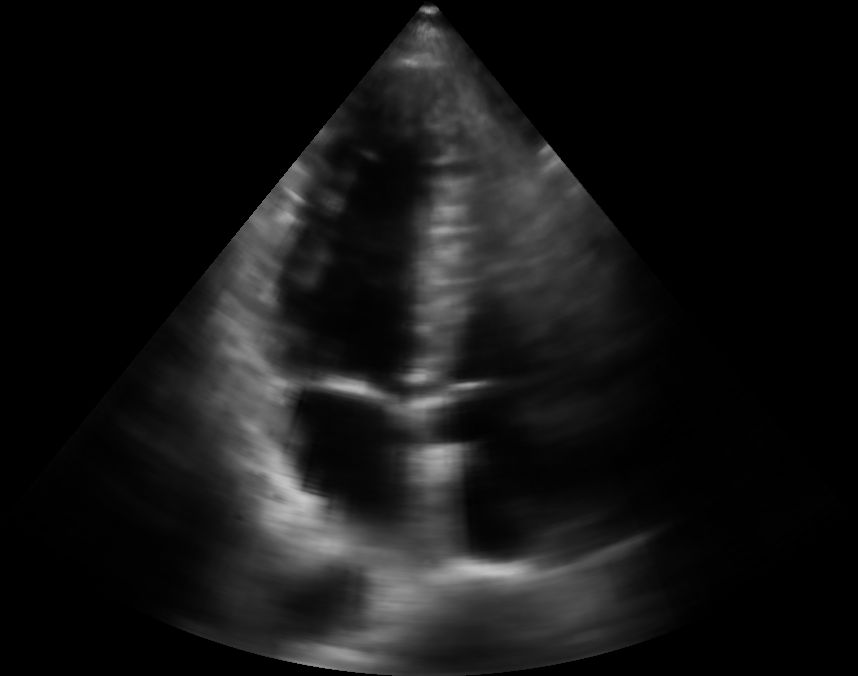} &
\includegraphics[width=0.20\textwidth]{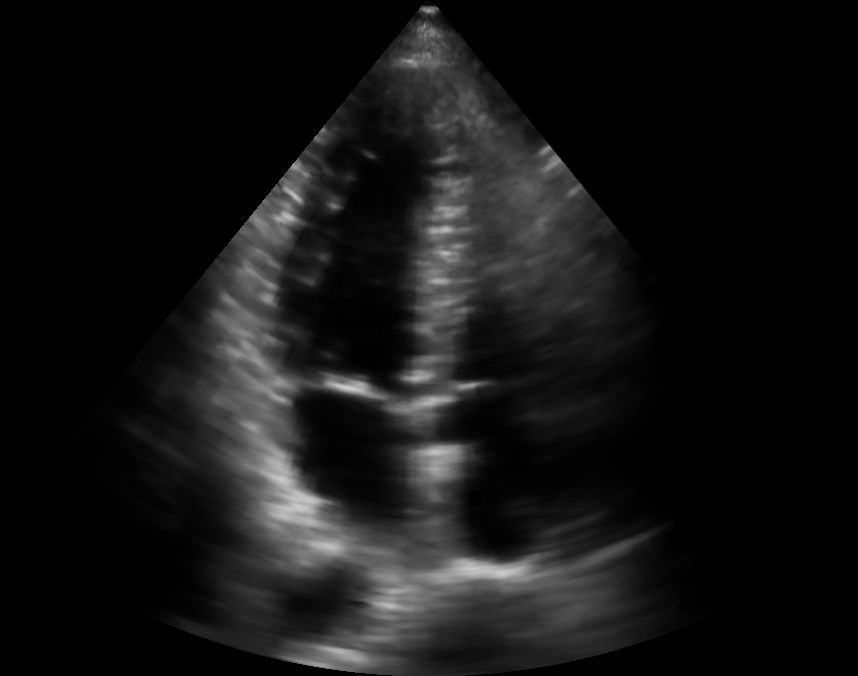}
\end{tabular}
\caption{Prediction results of the cardiac district, with the networks trained with 500 (a) and 1500 (b) images from the obstetric district, and 1500 images from multiple districts.\label{FIG:multidis2}}
\end{figure*}

Comparing the results on the other test data sets (e.g., cardiac Fig.~\ref{FIG:multidis2} and Table~\ref{TAB:multidis}), the network trained with images of multiple districts has better results than the networks trained with obstetric images only. The multiple district network better generalises on the denoising algorithm, more than on the anatomic features, thus generating fewer artefacts on the prediction. Furthermore, the multiple district network includes 500 cardiac images in its training data set, thus improving the prediction results on this district. As the main conclusion, a dedicated network for each anatomic district is the best solution for the prediction of the denoised ultrasound images of each specific district, if a sufficiently large data set is available for the training.
\subsection{Execution time and computational cost}\label{SEC:EXECTIME}
To test the training phase of the deep learning framework (Sect.~\ref{SEC:REQUS}) on the HPC framework (Sect.~\ref{SEC:HPC}), we exploit 8 nodes, each one composed of 32 cores and 4 accelerators, for a theoretical computational performance of 260 TFLOPS, and 220 GB of memory per node. The parallel implementation of the deep learning framework and the high hardware performance reduce the computation time of the training phase by at least 100 orders less than a serial implementation on a standard workstation.
\begin{table}[t]
\centering
\caption{With reference to the results in Figs.~\ref{FIG:multidis1},~\ref{FIG:multidis2}, we report the PSNR metric computed between the target and the prediction images, as average value among the 50 images of each test data set: obstetric (Ob.), muscle-skeletal (Msk.), cardiac (Card.), and multiple districts (Multi.). The network are trained with: single district (a, 500 obstetric images), single district (b, 1500 obstetric images) and multiple district images. \label{TAB:multidis}}
\begin{tabular}{c|cccc}
\textbf{Test data set} & Ob. & Msk. & Card. & Multi. \\ \hline
\textbf{Training with}	 & 		&		 &  &\\
Single district (a) & 35.93 & 25.88 & 28.46 & 29.47 \\
Single district (b) & 34.52 & 26.33 & 28.63 & 29.51 \\
Multiple district & 33.70 & 28.41 & 33.82 & 30.33
\end{tabular}
\end{table}

The execution time of the prediction is crucial for the real-time implementation of our framework. We test the denoising prediction on GPU-based hardware, which replicates the hardware of an ultrasound scanner currently in use. Given a set of ultrasound input images from different districts, the average execution time is 25 milliseconds; this result confirms that we achieve the real-time computation target, required by the industrial constraint.

We underline that the input resolution of the network is~$600 \times 600$, which is reached through the zero-padding of each input image. The computational cost of the prediction depends on the resolution of the input image and on the architecture of the network: in particular, the computational cost of a convolution operation is~$\mathcal{O} (r/s_r \cdot c/s_c)\cdot(f_r \cdot f_c)\cdot f$; in our application, the input image has a resolution of~$r=c=600$, the kernel-filter size on rows and columns is~$f_r = f_c = 4$, the stride on rows and columns is~$s_r = s_c = 2$, we use 10 convolution and 10 deconvolution operators, and a number of kernel-filters from 32 to 512.
\begin{figure*}
\centering
\begin{tabular}{cc}
(a) Input & (b) Target \\ 
\includegraphics[width=0.30\textwidth]{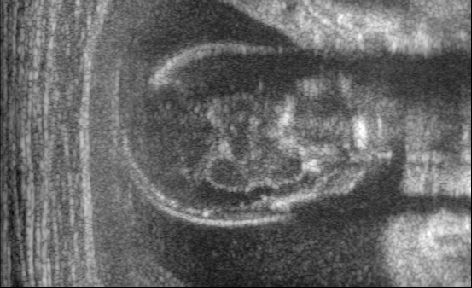} &
\includegraphics[width=0.30\textwidth]{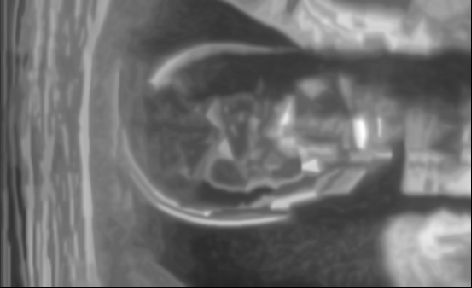} \\
Prediction: (c) Ours & (d) CNN\\
\includegraphics[width=0.30\textwidth]{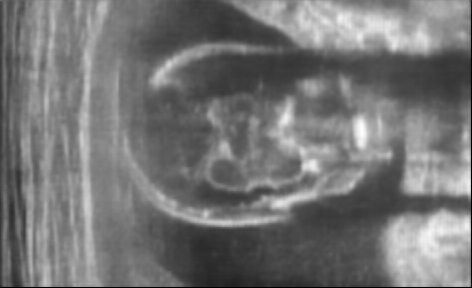} &
\includegraphics[width=0.30\textwidth]{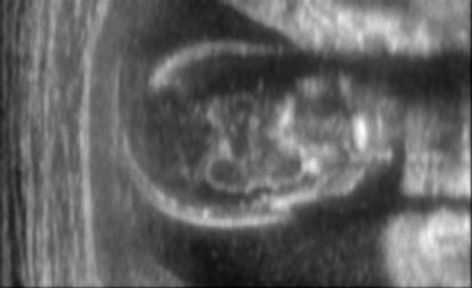}
\end{tabular}
\caption{(a) Input, (b) target, (c) our prediction based on Pix2Pix, and (d) CNN prediction for the obstetric district.\label{FIG:gancnncomparison}}
\end{figure*}
\subsection{Denosing on different learning architectures\label{SEC:NNcomparison}}
We compare the prediction results of two different networks: Pix2Pix and the (\emph{Matlab}) CNN, as part of our deep learning framework. Fig.~\ref{FIG:gancnncomparison} shows that Pix2Pix has better results than the CNN, in terms of blurring reduction, noise removal, and edge preservation. We also compare the quantitative metrics between the target images and the predicted images, on the test data set of 50 ultrasound images of the obstetric anatomic district (Sect.~\ref{SEC:DATASET4}). Pix2Pix has a PSNR average value of 36.07, and an SSIM average value of 0.878, while the CNN has a PSNR average value of 25.69, and an SSIM average value of 0.651. This result underlines that Pix2Pix outperforms the CNN as network architecture for our deep learning framework.

\section{Discussion\label{SEC:DISCUSSION}}
Several ultrasound machines manufactured by main competitors (e.g., Esaote, Philips) are equipped with GPU cards~\cite{urlesaote,urlphilips} Furthermore, some recent denoising methods for ultrasound images are developed on GPUs~\cite{palhano2011real,biswas2018ultrasound}, which are also used for denoising. Furthermore, the application of GPUs to image processing for future medical ultrasound imaging systems~\cite{so2011medical} presents the advantages of GPUs over CPUs in terms of performance, power consumption, and cost.

Denoising of ultrasound images is relevant both for post-processing and visual evaluation by medical experts. Despite some relevant works consider raw ultrasound images and videos for cardiac segmentation~\cite{liu2021deep,ouyang2020video}, several works show the benefits of denoising for \emph{segmentation}~\cite{yang2012shape,tang2010speckle,zhuang2019applicationMBEC}, \emph{feature extraction}~\cite{iakovidis2008fuzzy}, \emph{classification}~\cite{wei2020benign,sivanandan2021newMBEC}, \emph{super-resolution}~\cite{khavari2018non}, \emph{registration}~\cite{de20132d}, and \emph{texture analysis}~\cite{puri2005textureMBEC}. Furthermore, main ultrasound machine manufacturers include a denoising filter in their scanners~\cite{urlesaote,urlphilips2}.
In ultrasound denoising, the main goal is to achieve the best compromise between noise removal, features preservation, and real-time execution. The use of deep learning allows us to reach real-time performance, which is required by the clinical practice while preserving the denoising results of state-of-the-art methods, which are time-consuming. In our framework, deep learning preserves the quality of WNNM and reaches its results in real-time. Furthermore, the deep-learning for the real-time processing of ultrasound images has been applied in several works~\cite{liu2019deep,8808885}.

Our method reaches real-time performance and high-quality denoising results, through a learning-based approach. In contrast, fast handcrafted methods~\cite{garg2019despeckling} have lower results in terms of noise removal and edges enhancement; GPU-based methods~\cite{fredj2016real} have higher hardware requirements than our method; other denoising methods~\cite{xu2018trilateral} have good results in terms of noise removal, but they cannot reach a real-time implementation, due to high computational cost. Our framework also allows us to tune the denoising algorithm to obtain the best denoising results, as this tuning only affects the training phase, while the real-time computation of the denoised image is performed through the prediction of the network.

\section{Conclusions and future work\label{SEC:CONCLUSIONS}}
We have presented a novel deep learning framework for real-time denoising of ultrasound images, which is general enough to be applied to different anatomical districts and noise levels. As the main contribution, the proposed real-time denoising of ultrasound images is general in terms of the input data, i.e., type of noise (e.g., speckle, Gaussian noise), the resolution and the dimensionality of the input images (e.g., isotropic/anisotropic, 2D/3D images), the acquisition methodology, and the anatomical district. We also mention its generality in terms of building blocks and parameters of the deep learning framework, i.e., the denoising algorithms (e.g., WNNM, SAR-BM3D) and the deep learning architecture (e.g., Pix2Pix, VGG19).
 
As future work, we plan to apply our framework to data acquired with different methodologies (e.g., 3D ultrasound, MRI), also taking into account time-dependent data (e.g., ultrasound videos). Finally, the industrial and clinical validations of the proposed framework are under development, by comparing our results with tools currently used in medical clinics.
\paragraph{Acknowledgements} 
This research is carried out as part of an Industrial PhD project funded by CNR-IMATI and Esaote S.p.A. under the CNR-Confindustria agreement.
Tests on CINECA Cluster are supported by the ISCRA-C Project HP10CVHIXD.
\newpage
\begin{center}
\textbf{\large Additional material}
\end{center}
\section*{Quantitative comparison}
We compare the five selected denoising methods~(Sect.~\ref{sec:RESULTS}) on synthetic images, by adding speckle noise with different levels of noise intensity: given a noisy image \mbox{$Y = X + NX$}, where~$X$ is the normalised ground-truth image, we define the artificial multiplicative noise \mbox{$N(x) = \sqrt{12 \sigma} u$}, where \mbox{$u\sim\mathcal{U}(-0.5,0.5)$},~$\mathcal{U}$ is a uniform distribution,~$\sigma$ is the noise intensity, and~$x$ is a pixel of the image.

\begin{figure*}
\centering
\begin{tabular}{cccc}
Input image & Noisy image & WNNM & SAR-BM3D \\
\includegraphics[width=0.17\textwidth]{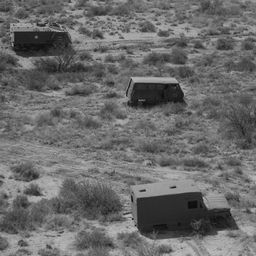} &
\includegraphics[width=0.17\textwidth]{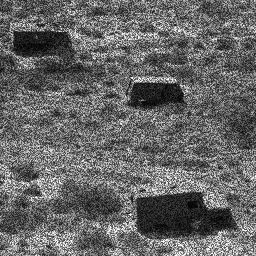} &
\includegraphics[width=0.17\textwidth]{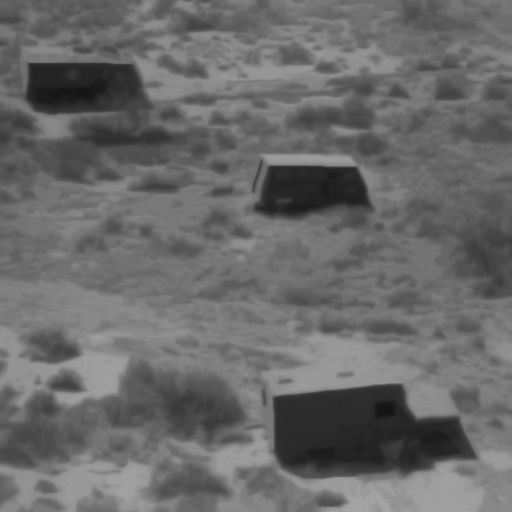} &
\includegraphics[width=0.17\textwidth]{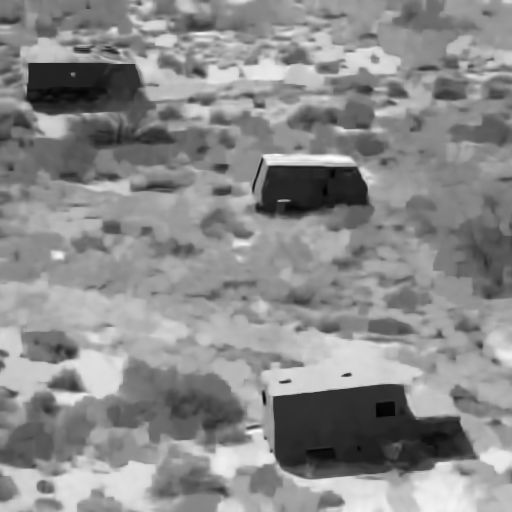} \\
PCA-BM3D & NCSR & BM-CNN & \\
\includegraphics[width=0.17\textwidth]{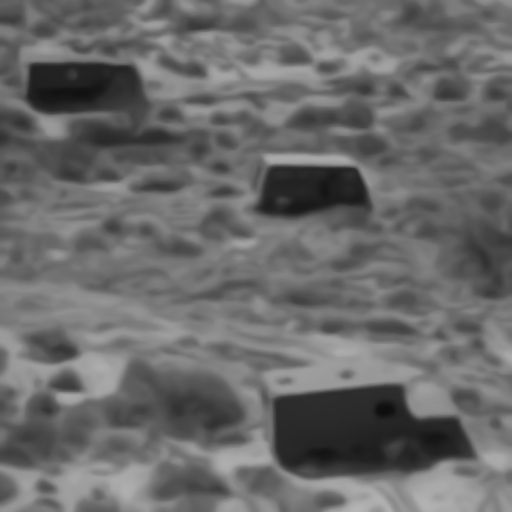} &
\includegraphics[width=0.17\textwidth]{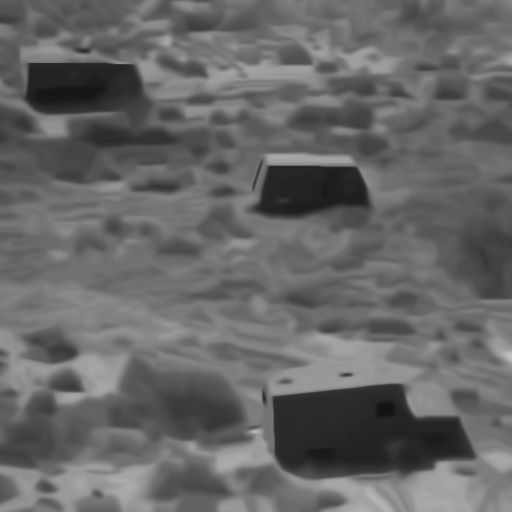} &
\includegraphics[width=0.17\textwidth]{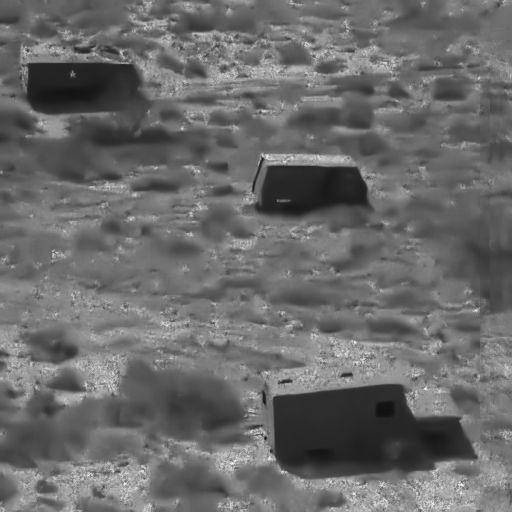}
\end{tabular}
\caption{Input (SIPI data set, van image), noisy (speckle noise intensity~$\sigma=0.10$), and denoised images. For error metrics, we refer the reader to Table~\ref{TAB:MISC}. \label{FIG:MISC}}
\end{figure*}

\begin{figure*}
\centering
\begin{tabular}{cccc}
Input image & Noisy image & WNNM & SAR-BM3D \\
\includegraphics[width=0.17\textwidth]{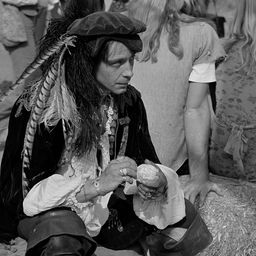} &
\includegraphics[width=0.17\textwidth]{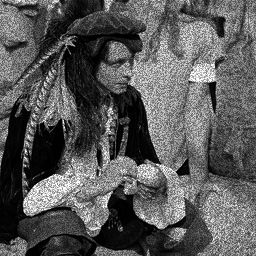} &
\includegraphics[width=0.17\textwidth]{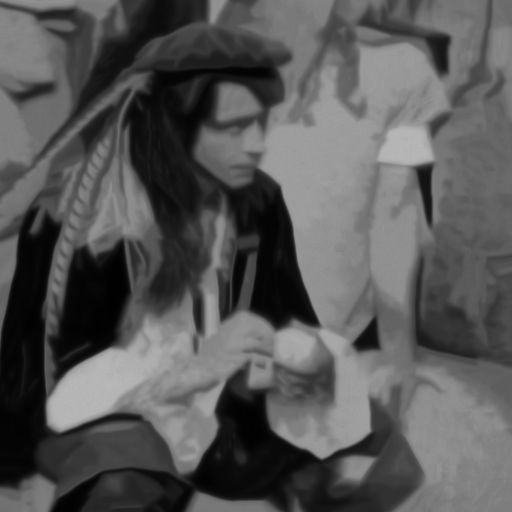} &
\includegraphics[width=0.17\textwidth]{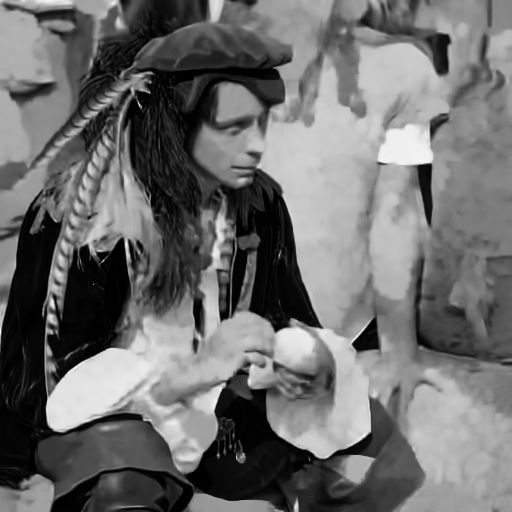} \\
PCA-BM3D & NCSR & BM-CNN & \\
\includegraphics[width=0.17\textwidth]{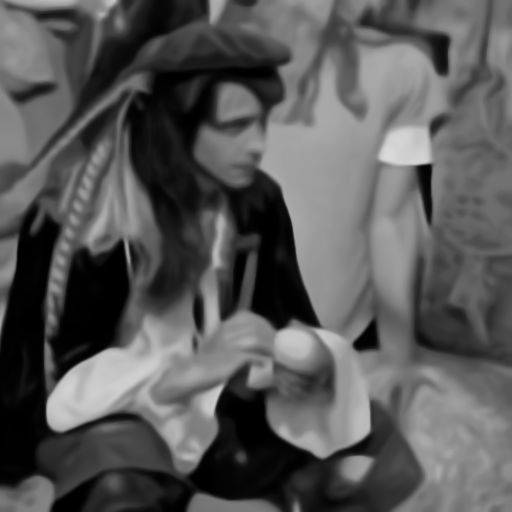} &
\includegraphics[width=0.17\textwidth]{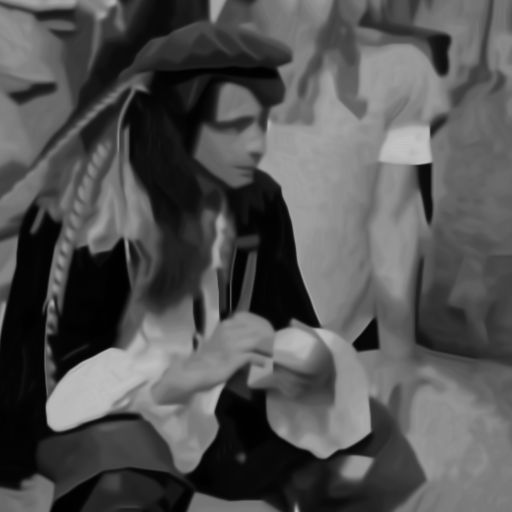} &
\includegraphics[width=0.17\textwidth]{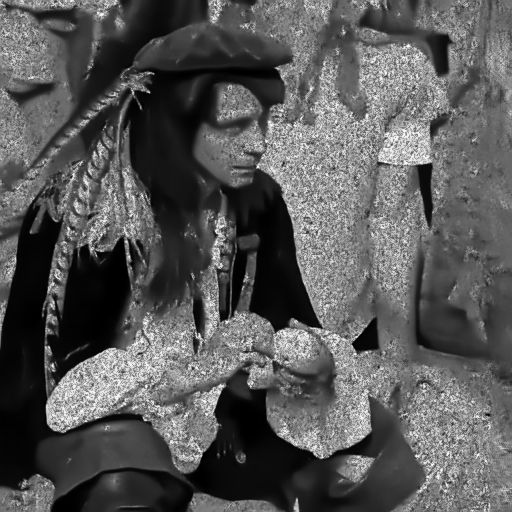}
\end{tabular}
\caption{Input (SIPI data set, man image), noisy (speckle noise intensity~$\sigma=0.20$), and denoised images. For error metrics, we refer the reader to Table~\ref{TAB:MISC}. \label{FIG:MISC2}}
\end{figure*}

The \emph{SIPI data set}~\cite{weber1997usc} is composed of 44 ground-truth images of different sizes, organised in different classes (e.g., humans, landscapes). We evaluate the efficiency of the denoising methods: WNNM has very good results in terms of noise removal, edge preservation (e.g., vehicles shape (Fig.~\ref{FIG:MISC}), and hat feathers (Fig.~\ref{FIG:MISC2}). SAR-BM3D has the best results in terms of noise removal; however, it does not correctly preserve the grey-scale values (e.g., boy's sleeve in Fig.~\ref{FIG:MISC2}) and it generates a blurred effect (e.g., grass and bushes in Fig.~\ref{FIG:MISC}). PCA-BM3D and NCSR show minor preservation of edges and details than WNNM (e.g., boy's face in Fig.~\ref{FIG:MISC2}). Finally, BM-CNN is not able to correctly remove the noise; this result underlines the importance of the training data set (e.g., the type and the intensity of the applied noise) when using a deep learning approach, and the necessity of using data-specific networks, instead of a generic-purpose one. 

Concerning the metrics introduced in Sect.~\ref{SEC:COMPARISON}, Table~\ref{TAB:MISC} summarises the results of the five denoising methods on the SIPI data set; we compute the average value of each metric (i.e., PSNR and SSIM) among 44 images of the data set, and report the average values when varying the intensity of the speckle noise. SAR-BM3D has the best results under these metrics, outperforming all the other methods. The NCSR, WNNM, and PCA-BM3D methods have good and similar results in terms of PSNR and SSIM indices. These four methods show a small degradation of the metrics values when increasing the noise intensity; this result is significant for ultrasound images, which generally have a different noise intensity, according to the anatomical district, the type of probe, and the data acquisition modality. Finally, BM-CNN shows a higher degradation of the PSNR and SSIM values, when increasing the noise intensity. The quantitative analysis is useful to compare methods with numerical measures, instead of performing only a visual evaluation. However, the main comparison among methods is the qualitative evaluation performed by the medical experts on ultrasound images, through the evaluation of the speckle noise removal and the preservation of anatomical features. We underline that, even if SAR-BM3D has better results than WNNM on synthetic images, WNNM has better performance on ultrasound images. Furthermore, our framework is general enough to use different denoising methods; two different learning networks can be trained, with WNNM and SAR-BM3D, to offer the physician the comparison between two denoising results.
\begin{table}[t]
\caption{PSNR and SSIM metrics of the denoising methods tested on the SIPI data set. For each~$\sigma$ value (i.e., the intensity of the speckle noise), we report the average metric computed on the 44 images of the data set. \label{TAB:MISC}}
\resizebox{\columnwidth}{!}{
\begin{tabular}{c|c|c|c|c|c|c|c|c}
\textbf{Metric }  & \multicolumn{4}{c|}{\textbf{PSNR}} & \multicolumn{4}{c}{\textbf{SSIM}} \\
\hline
Method~$\mid$~$\sigma$ & 0.05 & 0.1 & 0.2 & 0.3 & 0.05 & 0.1 & 0.2 & 0.3 \\ \hline
WNNM & 25.57 & 24.68 & 23.35 & 22.32 & 0.681 & 0.659 & 0.630 & 0.602 \\
SAR-BM3D &~$\mathbf{27.36}$ &~$\mathbf{26.01}$ &~$\mathbf{24.71}$ &~$\mathbf{23.65}$ & 0.730 &~$\mathbf{0.699}$ & ~$\mathbf{0.673}$ &~$\mathbf{0.651}$ \\
PCA-BM3D & 25.09 & 24.36 & 23.05 & 22.10 & 0.652 & 0.640 & 0.614 & 0.585 \\
NCSR & 26.60 & 25.36 & 23.61 & 22.36 & 0.665 & 0.669 & 0.619 & 0.588 \\
BM3D-CNN & 26.85 & 24.20 & 20.21 & 17.26 &~$\mathbf{0.733}$ & 0.569 & 0.31 & 0.215 
\end{tabular}}
\end{table}
\section*{Tuned-WNNM}
\begin{figure*}
\centering
\begin{tabular}{cccc}
Input image & Noisy image & WNNM & Tuned-WNNM \\
\includegraphics[width=0.17\textwidth]{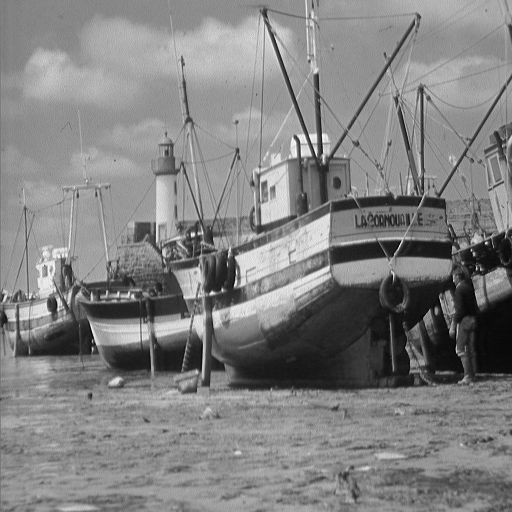} &
\includegraphics[width=0.17\textwidth]{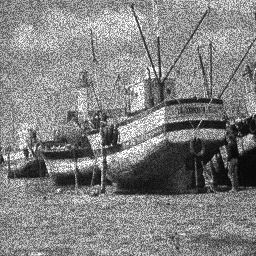} &
\includegraphics[width=0.17\textwidth]{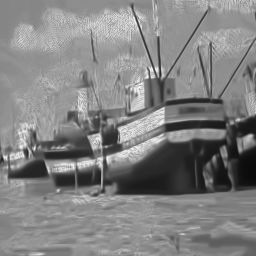} &
\includegraphics[width=0.17\textwidth]{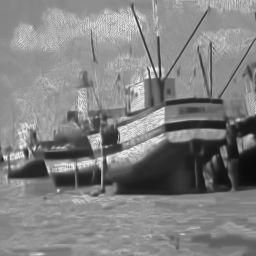}
\end{tabular}
\caption{Input ($256 \times 256$), noisy (speckle noise intensity~$\sigma=0.05$), and denoised images with WNNM and the tuned-WNNM. \label{FIG:TUNEDWNNM}}
\end{figure*}

Comparing the baseline WNNM with the tuned-WNNM on synthetic images, we improve the denoising quality (Fig.~\ref{FIG:TUNEDWNNM}) in terms of quantitative metrics; in fact, the output of WNNM has a PSNR value of 26.67, while the output of tuned-WNNM has a PSNR value of 26.74. Nevertheless, the execution time of WNNM is 94 seconds, while the execution time of tuned-WNNM is 260 seconds. We also compare tuned-WNNM and WNNM on the SIPI data set (Table~\ref{TAB:tuned2}). The aggregated results show that tuned-WNNM has slightly better performance with low noise intensity, while the results improve when the speckle noise is higher. For example, WNNM has an average PSNR value of 22.32 with a speckle noise of intensity~$\sigma=0.3$, while tuned-WNNM has a PSNR value of 22.61.
\begin{table}[t]
\caption{PSNR and SSIM metrics of WNNM and tuned-WNNM, tested on the SIPI data set. For each~$\sigma$ value (i.e., the intensity of the speckle noise), we report the average metric computed on the 44 images of the data set.
\label{TAB:tuned2}}
\resizebox{\columnwidth}{!}{
\begin{tabular}{c|c|c|c|c|c|c|c|c}
\textbf{Metric }  & \multicolumn{4}{c|}{\textbf{PSNR}} & \multicolumn{4}{c}{\textbf{SSIM}} \\
\hline
Method~$\mid$~$\sigma$ & 0.05 & 0.1 & 0.2 & 0.3 & 0.05 & 0.1 & 0.2 & 0.3 \\ \hline
WNNM & 25.57 & 24.68 & 23.35 & 22.32 & 0.681 & 0.659 & 0.630 & 0.602 \\
tuned-WNNM & ~$\mathbf{25.60}$ & ~$\mathbf{24.76}$ &~$\mathbf{23.49}$ &~$\mathbf{22.61}$ &~$\mathbf{0.685}$ &~$\mathbf{0.663}$ &~$\mathbf{0.642}$ & ~$\mathbf{0.627}$ \\
\end{tabular}}
\end{table}
\bibliographystyle{agsm} 
\bibliography{refs}
\end{document}